\documentclass[prx, twocolumn, superscriptaddress]{revtex4-2}
\usepackage{upgreek}
\usepackage{amsmath}
\usepackage{amssymb} 
\usepackage[utf8]{inputenc}
\usepackage{empheq}
\usepackage{placeins}
\usepackage[colorlinks=true,bookmarks=false]{hyperref}
\hypersetup{linkcolor=black,citecolor=black,urlcolor=blue}  
\usepackage{url}
\usepackage{setspace}

\makeatletter
\def\@ssect@ltx#1#2#3#4#5#6[#7]#8{%
  \def\H@svsec{\phantomsection}%
  \@tempskipa #5\relax
  \@ifdim{\@tempskipa>\z@}{%
    \begingroup
      \interlinepenalty \@M
      #6{%
       \@ifundefined{@hangfroms@#1}{\@hang@froms}{\csname @hangfroms@#1\endcsname}%
       {\hskip#3\relax\H@svsec}{#8}%
      }%
      \@@par
    \endgroup
    \@ifundefined{#1smark}{\@gobble}{\csname #1smark\endcsname}{#7}%
  }{%
    \def\@svsechd{%
      #6{%
       \@ifundefined{@runin@tos@#1}{\@runin@tos}{\csname @runin@tos@#1\endcsname}%
       {\hskip#3\relax\H@svsec}{#8}%
      }%
      \@ifundefined{#1smark}{\@gobble}{\csname #1smark\endcsname}{#7}%
      \addcontentsline{toc}{#1}{\protect\numberline{}#8}%
    }%
  }%
  \@xsect{#5}%
}%
\makeatother

\def\re{\mathrm{Re}}
\def\im{\mathrm{Im}}
\def\Re{\mathrm{Re}}
\def\Im{\mathrm{Im}}
\def\alphare{\alpha_{\re}}
\def\alphaim{\alpha_{\im}}
\newcommand{\sgn}[1]{\text{sgn}[{#1}]}
\newcommand{\bra}[1]{\ensuremath{\left\langle#1\right|}}
\newcommand{\ket}[1]{\ensuremath{\left|#1\right\rangle}}
\newcommand{\dv}[1]{\ensuremath{\frac{\text{d}}{\text{d}#1}}}

\newcommand{\beginsupplement}{%
        \setcounter{table}{0}
        \renewcommand{\thetable}{S\arabic{table}}%
        \setcounter{figure}{0}
        \setcounter{section}{0}
        \setcounter{equation}{0}
        \renewcommand{\thefigure}{S\arabic{figure}}
        \renewcommand{\theHfigure}{S\arabic{figure}}
        \renewcommand{\thesection}{S\Roman{section}}
        \renewcommand{\theHsection}{S\Roman{section}}
        \renewcommand{\theequation}{S\arabic{equation}}
		}%

\newcommand{\hiddensection}[1]{
    \stepcounter{section}
    \section*{\Roman{section}.\hspace{1em}{#1}}
}

\begin{document}
\title{Emerging dissipative phases in a superradiant quantum gas with tunable decay}

\author{Francesco Ferri}
\thanks{These authors contributed equally to this work.}
\affiliation{Institute for Quantum Electronics, ETH Zürich, 8093 Zürich, Switzerland}
\author{Rodrigo Rosa-Medina}
\thanks{These authors contributed equally to this work.}
\affiliation{Institute for Quantum Electronics, ETH Zürich, 8093 Zürich, Switzerland}
\author{Fabian Finger}
\affiliation{Institute for Quantum Electronics, ETH Zürich, 8093 Zürich, Switzerland}
\author{Nishant Dogra}
\thanks{Present address: Cavendish Laboratory, University of Cambridge, J. J. Thomson Avenue, Cambridge CB3 0HE, United Kingdom.}
\affiliation{Institute for Quantum Electronics, ETH Zürich, 8093 Zürich, Switzerland}
\author{Matteo Soriente}
\affiliation{Institute for Theoretical Physics, ETH Zürich, 8093 Zürich, Switzerland}
\author{Oded Zilberberg}
\affiliation{Institute for Theoretical Physics, ETH Zürich, 8093 Zürich, Switzerland}
\author{Tobias Donner}
\email{donner@phys.ethz.ch}
\affiliation{Institute for Quantum Electronics, ETH Zürich, 8093 Zürich, Switzerland}
\author{Tilman Esslinger}
\affiliation{Institute for Quantum Electronics, ETH Zürich, 8093 Zürich, Switzerland}
\date{\today}

\begin{abstract}
Exposing a many-body system to external drives and losses can transform
the nature of its phases and opens perspectives for engineering
new properties of matter. How such characteristics are
related to the underlying microscopic processes of the driven and
dissipative system is a fundamental question. Here we address this point in
a quantum gas that is strongly coupled to a lossy optical cavity mode using
two independent Raman drives, which act on the spin and motional degrees of
freedom of the atoms. This setting allows us to control the competition
between coherent dynamics and dissipation by adjusting the imbalance
between the drives. For strong enough coupling, the transition to a
superradiant phase occurs, as is the case for a closed system. Yet, by
imbalancing the drives we can enter a dissipation-stabilized normal phase
and a region of multistability. Measuring the properties of excitations
on top of the out-of-equilibrium phases
reveals the microscopic elementary
processes in the open system. Our findings provide prospects for studying
squeezing in non-Hermitian systems, quantum jumps in superradiance, and
dynamical spin-orbit coupling in a dissipative setting.
\end{abstract}
\maketitle

\hiddensection{Introduction}
Open many-body systems can annul fundamental laws that typically govern the physics of systems in thermal equilibrium. In the idealized case of an ensemble of interacting particles, isolated from the environment and at zero temperature, the ground state is set by energy minimization and phase transitions arise from competing energy contributions~\cite{Sachdev_2011,Carr2010Understanding}. In open systems however, the interplay between coherent dynamics within the system and its interaction with the environment gives rise to a much richer set of phenomena~\cite{Muller_2012, Daley_2014, Carusotto2013Quantum,Sieberer_2016,Soriente_2021}. Such interaction is not only unavoidable, but can be exploited via the engineering of external drives and coupling to specific baths~\cite{Diehl_2011, Blatt_2011, Wineland_2013, Polzik_2011}.
The experimental access to many-body ground states provided by ultra-cold atoms~\cite{Bloch2008Many,Lewenstein2012Ultracold, Schmiedmayer_2015} laid the foundation for a recent revival of interest in many-body systems interacting with their environment. Experimental observations that are specifically due to the system’s openness include the emergence of bistability~\cite{Ott_2016, Ott_2017, Houck_2017, Imamoglu_2018}, the stabilization of insulating phases~\cite{Tomita_2017, Simon_2019}, the access to absorbing-states phase transitions~\cite{Lesanovsky_2018}, the appearance of dissipation-induced instabilities~\cite{Dogra_2019} and time crystals~ \cite{Hemmerich_2021}, and the change in correlation properties~\cite{Gorshkov_2020, Porto_2020} that can signal non-Hermitian phase transitions~\cite{Weitz_2020}. 

Besides their fundamental interest, non-equilibrium phenomena bear the prospect of becoming a powerful tool for engineering new materials ranging from exciton condensates to light-induced superconductors~\cite{Basov_2017, Byrnes_2014, Cavalleri_2018, Imamoglu_2014, Orgiu_2015}. The properties of these phases of matter emerge from tuning the elementary excitations by hybridization with the light field~\cite{Georges_2019, Demler_2020_2}, which provides a natural coupling to the external environment in presence of optical drives and losses~\cite{Galitski_2019, Jaksch_2020}.
To gain further insight into this phenomenology, it is desirable to achieve good experimental control over coherent and dissipative channels and at the same time to gain access to the microscopic properties lying at the origin of the macroscopic phases~\cite{Carusotto_2020}.

\begin{figure}[th]
    \centering
    \includegraphics[width=\columnwidth]{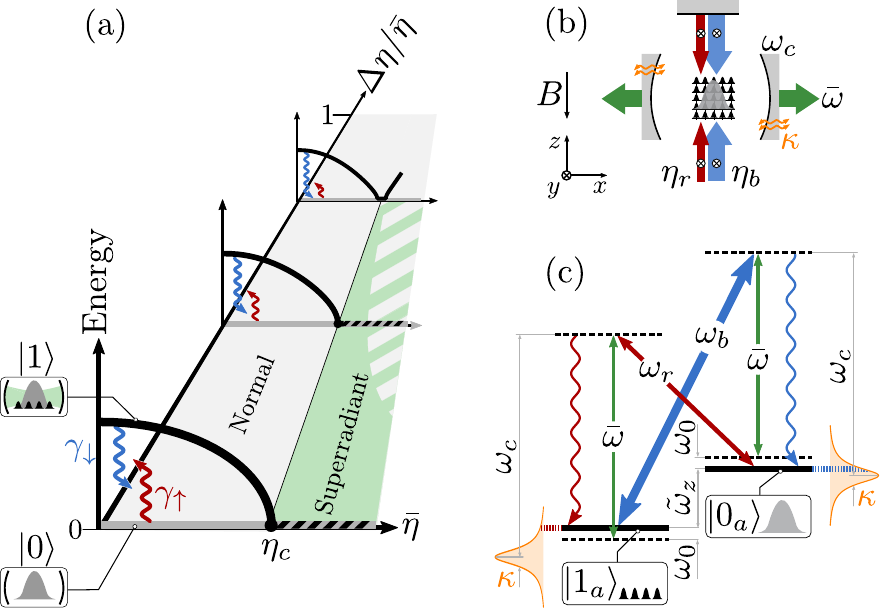}
    \caption{\textbf{Competing coherent coupling and dissipation at a superradiant phase transition.} (a) A quantum gas interacts coherently with a cavity mode via two tunable drives with mean coupling strength $\bar{\eta}$ and imbalance $\Delta\eta$, giving rise to two low-energy polariton modes $\ket{0}$ and $\ket{1}$, corresponding to decoupled and coupled light-matter modes, respectively. Increasing $\bar{\eta}$ softens the energy of $\ket{1}$ (black line); cavity dissipation is responsible for the effective damping ($\gamma_\downarrow$) and amplification ($\gamma_\uparrow$) of the soft-mode polariton $\ket{1}$. For small enough $\Delta\eta/\bar{\eta}$, the rates  $\gamma_\downarrow$, $\gamma_\uparrow$ are balanced, and the mode softening to zero energy at $\bar{\eta}=\eta_c$ is accompanied by a phase transition from the normal phase populating only $\ket{0}$ (gray shade) to the superradiant phase (green shade), where $\ket{1}$ is occupied. By increasing  $\Delta\eta/\bar{\eta}$, the dominating damping $\gamma_\downarrow$ of the soft mode leads first to bistability (green-gray hashed region), and finally to the suppression of the superradiant transition. (b) Sketch of the experimental setup and (c) corresponding level scheme. A BEC inside a high-finesse cavity (with resonant frequency $\omega_c$ and field decay rate $\kappa$) is illuminated transversally by two Raman lasers with coupling strengths $\eta_{b(r)}$ and frequencies $\omega_{b(r)}$. The BEC ($\ket{0_a}$) couples to a spatially-modulated state $\ket{1_a}$ of the neighbouring Zeeman sublevel, separated by an energy $\hbar\tilde{\omega}_z=\hbar(\omega_z-2\omega_\mathrm{rec})$, with Zeeman splitting $\hbar\omega_z$ and recoil energy $\hbar\omega_\mathrm{rec}$. In the superradiant phase, a coherent field at frequency $\bar{\omega}$ (green) builds up in the cavity. The two-photon transitions driven by each pump and the cavity field at $\bar{\omega}$ are detuned from the bare atomic states $\ket{0_a}$, $\ket{1_a}$ by $\mp\omega_0$, as indicated by the lower dashed lines. The dissipative channels between modes $\ket{0}$ and $\ket{1}$ shown in (a) are due to Raman scattering of photons from each single drive into the cavity (blue and red wiggly arrows), and subsequent photon loss. The Lorentzian density of states of the cavity is sketched in orange.}
        \label{fig:Fig1}
\end{figure}

In this work, we engineer a driven-dissipative many-body system with global-range interactions, that is subject to tunable coherent and dissipative channels. Our realization employs a quantum gas strongly coupled to an optical cavity~\cite{Ritsch2013Cold,Mivehvar_2021}. Building on schemes that have been extensively exploited with thermal atoms where the atomic spin is coupled to light fields~\cite{Polzik_2010, Vuletic_2007, Barrett_2017, SchleierSmith_2019, Thompson_2018, Thompson_2020}, our implementation also involves the density degree of freedom of the gas~\cite{Lev_2018, StamperKurn_2018, Landini_2018}. We employ two Raman laser drives to control the strength of the co- and counter-rotating terms of the resulting light-matter coupling independently. In combination with photon loss from the cavity, this allows us to explore different regimes of competing coherent coupling and dissipation. Schematically, one can identify the following processes (Fig.~\ref{fig:Fig1}): the combined action of the two drives coherently mixes two many-body atomic states ($\ket{0_a}$, $\ket{1_a}$) with the cavity field into polariton modes. As the strength of the drives increases, the excited polariton mode $\ket{1}$ softens to the lowest-energy one $\ket{0}$, and a second-order phase transition occurs from a normal to a superradiant phase that is phase-locked to the drives. In addition to this coherent process, each individual drive can induce transitions from one atomic state to the other ($\ket{0_a}\leftrightarrows\ket{1_a}$), where photons are scattered into the cavity mode and successively lost. As a result, adjusting the relative strength of the two lasers yields to tuning the effective polariton dissipation. This leads to qualitative changes in the phase diagram of the system, with the appearance of a dissipation-stabilized phase and a discontinuous phase transition in a multistable region. We relate these phenomena to the changing properties of the polaritonic excitations, which we characterize experimentally and theoretically.

\hiddensection{Setup and tunable decay}
In our experiments, we trap a Bose-Einstein condensate (BEC) inside a high-finesse optical cavity and pump the atoms with a two-frequency optical standing wave, perpendicular to the cavity axis [Fig.~\ref{fig:Fig1}(b)]. The BEC is formed by $N=10^5$ atoms of $^{87}$Rb, prepared in the $m_F=-1$ sublevel of the $F=1$ ground state hyperfine manifold. A magnetic field along the $z$-direction is applied to generate a Zeeman splitting $\omega_z=2\pi\cdot 48~$MHz between the initial state and the $m_F=0$ sublevel. The driving fields are far red-detuned from the atomic resonance with frequencies $\omega_b,\,\omega_r$ that lie on opposite sides of the dispersively shifted cavity resonance $\omega_c$, i.e., $\omega_r<\omega_c<\omega_b$, and $\omega_b-\omega_r\sim2\omega_z$. The standing-wave modulations of the two drives overlap at the position of the atoms, forming a one-dimensional lattice potential with spacing $\lambda/2=784.7/2~$nm. Each pump beam realizes a cavity-assisted Raman coupling between the $m_F=-1$ and $m_F=0$ levels, as sketched in Fig.~\ref{fig:Fig1}(c). The resulting system is effectively described using two atomic modes: the initial ground state of the BEC $\ket{0_a}$, and the excited-momentum state of the neighbouring Zeeman sublevel $\ket{1_a}\propto\cos(k_\mathrm{rec}\hat{x})\cos(k_\mathrm{rec}\hat{z})\hat{F}_{+}\ket{0_a}$, with $\hat{F}_{+}$ being the raising spin-operator in the $F=1$ manifold, and $\hbar k_{\mathrm{rec}}=2\pi\hbar/\lambda$ the recoil momentum. When increasing the driving strength, the ground state $\ket{0_a}$ evolves from a harmonically confined BEC to a stack of pancake-shaped BECs trapped in the maxima of the standing-wave drives.

In a rotating frame defined by the driving frequencies, the coherent dynamics of our system is described by the many-body Hamiltonian
\begin{equation}
\hat{H} = -\hbar\Delta_c \hat{a}^\dagger \hat{a} + \hbar\omega_0 \hat{J}_z + 2\hbar\bar{\eta}(\hat{a}+\hat{a}^\dagger)\hat{J}_x + 2i\hbar\Delta\eta(\hat{a}-\hat{a}^\dagger)\hat{J}_y,
\label{eq:Hamiltonian}
\end{equation}
as detailed in~\cite{SI}. Here, $\hat{J}_{i=x,y,z}$ are the components of the pseudo-spin operator associated with the two many-body states $\ket{0_a}$ ($\langle \hat{J}_z \rangle=-N/2$) and $\ket{1_a}$ ($\langle \hat{J}_z\rangle=N/2$), and $\hat{a}~(\hat{a}^\dagger)$ is the annihilation (creation) operator of the relevant cavity mode. The effective atomic frequency is $\omega_0=(\omega_b-\omega_r)/2-\omega_z+2\omega_{\mathrm{rec}}$, with the recoil energy $\hbar\omega_{\mathrm{rec}}$, and the cavity detuning $\Delta_c=\bar{\omega}-\omega_c$, where $\bar{\omega}=(\omega_b+\omega_r)/2$ is the mean of the pump frequencies. The light-matter coupling strengths are parametrized by $\bar{\eta}=(\eta_b+\eta_r)/2$ and $\Delta\eta=(\eta_b-\eta_r)/2$, with the cavity-assisted Raman coupling $\eta_{b(r)}$ arising from the $\omega_{b(r)}$ pump and the cavity mode. These two-photon Raman couplings implement the independently tunable co- and counter-rotating terms of the light-matter interaction. The dynamics of the open system due to photon losses is well described by a master equation $\dot{\hat{\rho}}=-i/\hbar[\hat{H}, \hat{\rho}]+\hat{\mathcal{L}}[\hat{\rho}]$, where the Lindblad superoperator $\hat{\mathcal{L}}[\hat{\rho}]=\kappa\left[2\hat{a}\hat{\rho}\hat{a}^\dagger-\{\hat{a}^\dagger \hat{a},\hat{\rho}\}\right]$ accounts for the cavity field decay at rate $\kappa=2\pi\cdot 1.25~$MHz. The model introduced here is a generalized Dicke model that is predicted to exhibit rich phenomenology, see Ref.~\cite{Dalla_Torre_2019} and the more recent Refs.~\cite{Parkins_2020, Demler_2020, Soriente_2020}. Correspondingly, first experiments exploring effective Dicke models with tunable co-and counter-rotating terms and different beam geometries have been carried out using thermal atoms~\cite{Barrett_2017, Barrett_2018}, whose motional state is not well defined.

When the co- and counter-rotating coupling terms are balanced ($\Delta\eta=0$), Eq.~\eqref{eq:Hamiltonian} reduces to the Dicke Hamiltonian~\cite{Dicke_1954, Dimer_2007}. In this limit, at low coupling $\bar{\eta}$, the system is in the normal phase with $\langle \hat{J}_x\rangle=0$, $\langle \hat{a}\rangle=0$, and the lowest-energy polariton mode $\ket{0}$ is mostly occupied. By increasing the coupling $\bar{\eta}$, the energy of the first excited polariton $\ket{1}$, admixing both atomic states $\ket{0_a}$, $\ket{1_a}$ and the cavity photons, softens. As the energy of the polariton $\ket{1}$ reaches zero, the system undergoes a second-order phase transition to the self-organized superradiant phase with $\langle \hat{J}_x\rangle\neq 0$, $\langle \hat{a}\rangle\neq 0$ [Fig.~\ref{fig:Fig1}(a)]~\cite{Baumann_2010, Lev_2018, Mivehvar_2021}. The transition occurs at a collectively-enhanced critical coupling $\eta_c\sqrt{N}=\sqrt{-\omega_0(\Delta_c^2+\kappa^2)/(4\Delta_c)}$, which is only slightly shifted from the closed-system critical point for our parameters~\cite{SI}.
On the other hand, if an imbalance $\Delta\eta$ between the coupling of co- and counter-rotating terms is introduced, the effect of dissipation on the system becomes qualitatively different. Specifically, due to cavity loss, each Raman drive $\eta_{b(r)}$ is responsible for an effective decay $\gamma_{\downarrow(\uparrow)}$ of the polariton mode $\ket{1}(\ket{0})$ towards mode $\ket{0}(\ket{1})$. In the parameter regime of our experiment $\omega_0\ll\kappa$, the effective decay rates take the form
\begin{equation}
\gamma_{\downarrow(\uparrow)}=N\frac{\kappa}{\Delta_c^2+\kappa^2}\eta_{b(r)}^2,
\label{eq:damping}
\end{equation}
which we derive using an effective Keldysh action for the polariton modes~\cite{SI}. We identify that the microscopic process corresponding to the effective decay $\gamma_{\downarrow(\uparrow)}$ is a collectively-enhanced Raman scattering driven by the $\omega_{b(r)}$ pump beam, into the dissipatively-broadened density of states of the cavity, as sketched in Fig.~\ref{fig:Fig1}(c). This mechanism is analogous to the Raman decay lying at the heart of superradiant Raman lasers~\cite{Kasevich_2011, Thompson_2012, Thompson_2014, Molmer_2018}. Note that the effective decays~\eqref{eq:damping} are independent of the phase of the cavity field. This is different from the coherent Raman couplings leading to the superradiant phase, where the intra-cavity field is always in- or out-of-phase with the effective driving field $\bar{\omega}$~\cite{Baumann_2011}.
The two processes $\gamma_{\downarrow(\uparrow)}$ act against each other, either damping or amplifying the population of mode $\ket{1}$, such that global dissipative effects on the system are enhanced when the pumps are not balanced. In particular, as we experimentally demonstrate, the effective damping generated by these processes leads to a dramatic modification of the superradiant phase transition, as well as to new regions of multistability and hysteresis.\\

\hiddensection{Phase diagram}
\label{sec:Phase diagram}
\begin{figure}[thbp]
    \centering
    \includegraphics[width=0.85\columnwidth]{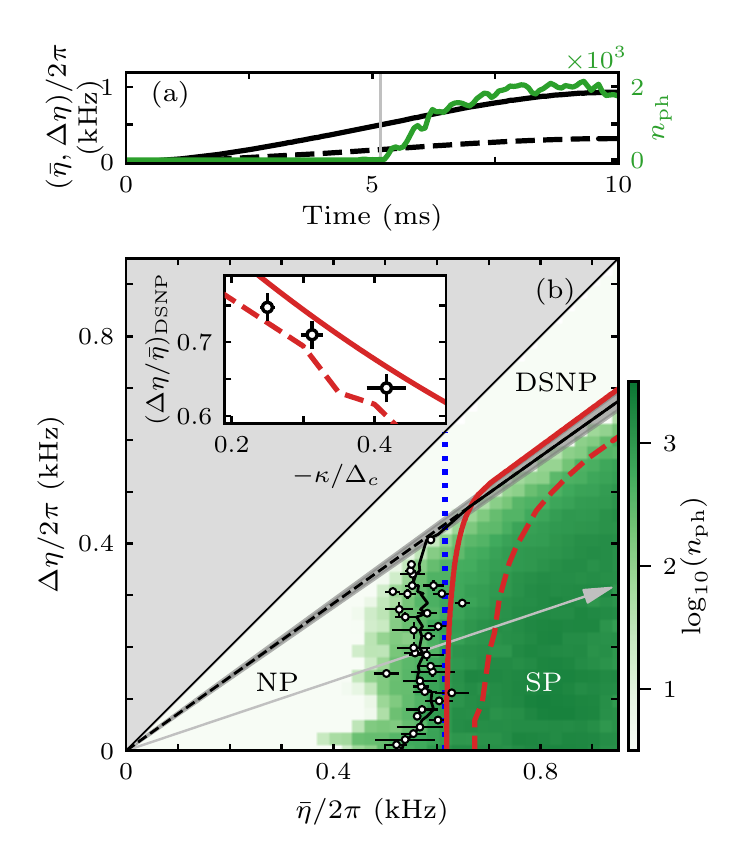}
    \caption{\textbf{Phase diagram}. (a) Experimental protocol. We ramp up the couplings $\bar{\eta}$ (solid black), $\Delta\eta$ (dashed black) within $10~$ms at constant ratio $\Delta\eta/\bar{\eta}$. An exemplary time trace of the mean intra-cavity photon number $n_\mathrm{ph}$ for $\Delta\eta/\bar{\eta}=0.34$ is shown (green). For any $\Delta\eta/\bar{\eta}<0.71(2)$, a coherent cavity field builds up in the cavity (superradiant phase) above a critical coupling (gray vertical line). 
    (b) Green: intra-cavity photon number $n_\mathrm{ph}$ as a function of the couplings $\bar{\eta},\Delta\eta$, obtained by implementing the protocol shown in (a) for $51$ different values of $\Delta\eta/\bar{\eta}$, and averaging over $5$ repetitions. 
Labels NP, SP and DSNP stand for normal, superradiant and dissipation-stabilized normal phase, respectively. The dots indicate the critical point extracted from the measured photon traces, and averaged within each subset of constant $\Delta\eta/\bar{\eta}$. The slope $(\Delta\eta/\bar{\eta})_\textrm{DSNP}$ of the dashed line corresponds to the smallest value of $\Delta\eta/\bar{\eta}$ at which the SP is not present (the uncertainty on the slope is marked as a shaded gray region around the line). The black line is a guide to the eye through the critical points (obtained as a sliding average over 7 points) and along the measured boundary of the DSNP. We mark the boundary of the SP obtained from a mean-field stability analysis (solid, red), and from a numerical simulation of the measurement protocol (dashed, red)~\cite{Soriente_2018, SI}. For comparison, the analytical mean-field result for the closed system is also plotted (blue, dotted). The arrow indicates the measurement path followed in (a). Inset: slope $(\Delta\eta/\bar{\eta})_\textrm{DSNP}$  extracted from phase diagrams measured at different cavity detunings $\Delta_c$, and plotted vs. $-\kappa/\Delta_c$. The lines are predictions from analytical (solid) and numerical (dashed) calculations. 
For this measurement, $N=1.28(8)\times10^5$, $\omega_0=2\pi\cdot44(2)~$kHz and, in (a), (b), $\Delta_c=-2\pi\cdot4.0(2)~$MHz.}
        \label{fig:Fig2}
\end{figure}

We restrict the experiments to the parameter space $0\leq\Delta\eta\leq\bar{\eta}$; the properties of the system in the region $0\leq\bar{\eta}\leq\Delta\eta$ are mirrored, cf.~Eq.~\eqref{eq:Hamiltonian}. We map out the phase diagram of the system by ramping up the power of the pump beams while keeping the ratio $\Delta\eta/\bar{\eta}$ constant, and monitoring the cavity output with a heterodyne detection, see Fig.~\ref{fig:Fig2}(a). The onset of a superradiant phase is signalled by the build-up of a coherent cavity field with frequency $\bar{\omega}$ above a critical coupling strength. 
We show in Fig.~\ref{fig:Fig2}(b) the measured mean intra-cavity photon number $n_{\mathrm{ph}}$ in the $(\bar{\eta},\Delta\eta)$ parameter space. 
At small imbalances $\Delta\eta\ll\bar{\eta}$, the phenomenology of the Dicke phase transition is observed, and the value of the coupling $\bar{\eta}\approx\eta_c$ at which the transition occurs depends only weakly on $\Delta\eta$. In contrast, at larger ratios $\Delta\eta/\bar{\eta}> 0.71(2)$, the superradiant phase transition is suppressed and the system remains in the normal phase at values of $\bar{\eta}$ largely above $\eta_c$.  We compare the measured critical couplings with the phase boundaries obtained from a mean-field treatment of our driven-dissipative model, and observe consistency between the experiment and our theoretical description. The phase boundaries are calculated from both the steady-state solutions and numerical simulations including time-varying couplings, with no free parameters (see \cite{SI} for details).

The existence of a dissipation-stabilized normal phase near the line $\Delta\eta/\bar{\eta}=1$ is a consequence of the open character of our system, as pointed out in previous theoretical works~\cite{Soriente_2018, Parkins_2020}. To further characterize this phase, we measured the full phase diagram of the system for different cavity detunings  $\Delta_c$~\cite{SI}. We observe that the width of the dissipation-stabilized normal phase increases for smaller detunings [inset of Fig.~\ref{fig:Fig2}(c)]. This agrees with the predictions of our theoretical model, by which we find that the slope $(\Delta\eta/\bar{\eta})_\mathrm{DSNP}$ of the boundary between the dissipation-stabilized normal phase and the superradiant phase at large couplings $\bar{\eta}\gg\eta_c$ is given by $(\Delta\eta/\bar{\eta})_\mathrm{DSNP}=\kappa/\Delta_c\left(1-\sqrt{1+\Delta_c^2/\kappa^2}\right)$~\cite{SI}.\\

\begin{figure}
    \centering
    \includegraphics[width=1\columnwidth]{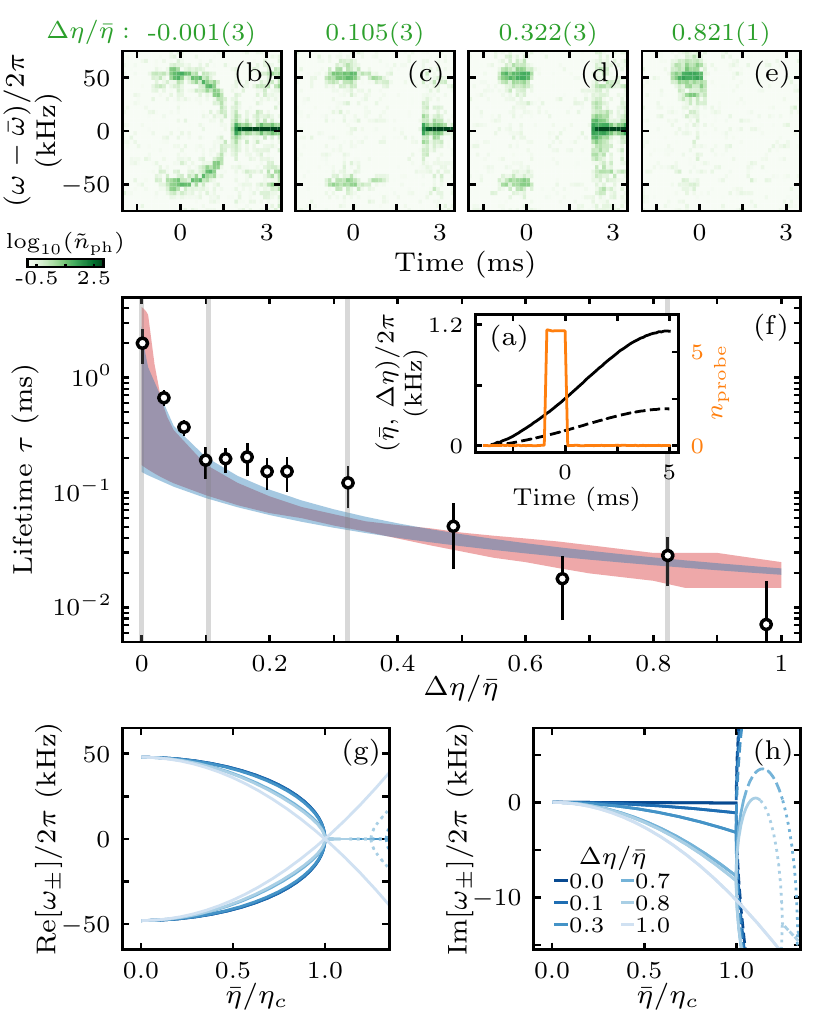}
    \caption{\textbf{Properties of the excited polariton.} (a) Experimental protocol. While ramping up the couplings $\bar{\eta}$ (solid black) and $\Delta\eta$ (dashed) at constant $\Delta\eta/\bar{\eta}$, we populate the excited mode $\ket{1}$ using a weak excitation pulse along the cavity axis (orange). Throughout this measurement, the maximal coupling $\bar{\eta}$ at the end of the ramp is $2\pi\cdot 1.16(5)~$kHz. (b-e) Exemplary heterodyne spectrograms for different ratios $\Delta\eta/\bar{\eta}$, showing the frequency-resolved mean number of photons $\tilde{n}_{\mathrm{ph}}$, as a function of frequency and time~\cite{SI}. In (e), the large imbalance $\Delta\eta/\eta$ suppresses the superradiant phase transition. (f) Data points: measured lifetime of the induced excitations as a function of $\Delta\eta/\bar{\eta}$~\cite{SI}. Blue shaded region: predicted lifetime from the analytical eigenvalues [cf. (g) and (h)], assuming a phenomenological atomic dephasing between $0~$Hz (upper bound) and $2\pi\cdot 500~$Hz (lower bound); the latter is of the order of the $s$-wave scattering rate. Red shaded region: numerical simulation results. The values of $\Delta\eta/\bar{\eta}$ shown in (b-e) are marked with gray vertical lines. (g,h) Excitation eigenvalues of the system, linearized around the normal phase, and assuming zero atomic dephasing. Colors indicate different $\Delta\eta/\bar{\eta}$ values, line shape indicates stable normal phase (solid), stable superradiant phase (dashed), bistability (dotted). For the results presented here, $N=9.6(4)\times10^4$, $\omega_0=2\pi\cdot48(4)~$kHz, $\Delta_c=-2\pi\cdot5.8(1)~$MHz, $\delta_{\mathrm{probe}}=2\pi\cdot 2.0(4)~$kHz.}
        \label{fig:Fig3}
\end{figure}

\hiddensection{Probing the excitation spectrum}
\label{sec:Excitations}
The effective damping induced by the relative coupling imbalance $\Delta\eta/\bar{\eta}$ leads to a profound change in the superradiant phase transition and even suppresses it. This observation is closely linked to the system's spectrum of collective excitations. We implement an experimental protocol that allows for non-destructive, real-time monitoring of the free evolution of the excited polariton mode $\ket{1}$, both in amplitude and in frequency. We promote a small population of mode $\ket{0}$ to mode $\ket{1}$ by means of a Bragg scattering process involving the transverse pump beams and a $1~$ms-long laser pulse injected into the cavity at frequency $\bar{\omega}+\omega_0+\delta_{\mathrm{probe}}$. This small occupancy of the excited mode produces a scattering of a weak field from the pumps into the cavity also after the pulse ended, which we monitor with the frequency-resolving heterodyne detector~\cite{Brennecke_2013}. We shine the excitation pulse at a fixed time while ramping up both couplings at constant ratios $\Delta\eta/\bar{\eta}$. At the falling edge of the pulse, the mean pump intensities correspond to $\bar{\eta}/\eta_c\approx0.6$ [Fig.~\ref{fig:Fig3}(a)].

In Figs.~\ref{fig:Fig3}(b-e), we present exemplary spectrograms of the cavity field showing the excitation pulse and the subsequent evolution of the system for increasingly larger values of $\Delta\eta/\bar{\eta}$. In the Dicke limit [$\Delta\eta/\bar{\eta}=0$, Fig.~\ref{fig:Fig3}(b)], the main components of the spectrum evolve towards $\bar{\omega}$ as the coupling $\bar{\eta}$ is swept to larger values, reflecting the softening of the excited mode energy. At the critical point $\bar{\eta}=\eta_c$, the energy gap between mode $\ket{0}$ and the soft mode $\ket{1}$ vanishes, and the superradiant phase transition occurs, signalled by the build-up of a strong coherent field at frequency $\bar{\omega}$. As the relative imbalance $\Delta\eta/\bar{\eta}$ is increased, the mode softening is accompanied by a faster decay of the excitation amplitude during the experiment. The $1/e$-lifetime $\tau$ of the free excitation extracted from the integrated spectrograms decreases rapidly as the relative imbalance $\Delta\eta/\bar{\eta}$ is increased [Fig.~\ref{fig:Fig3}(f)]. At the same time, the superradiant phase transition occurs at a coupling $\bar{\eta}$ that depends only weakly on the coupling imbalance $\Delta\eta$ until, for large enough $\Delta\eta/\bar{\eta}$, the transition is fully suppressed. 

The connection between the damping of the excitations and the suppression of the superradiant phase transition can be understood by analyzing the excitation spectrum of the open system. We linearize the mean-field equations of motion around the normal phase and study the low-energy eigenfrequencies $\omega_{\pm}$ of the corresponding dynamical matrix, as a function of the coupling $\bar{\eta}/\eta_c$, and for different values of $\Delta\eta/\bar{\eta}$ [Fig.~\ref{fig:Fig3}(g,h)]. The real part of the spectrum captures the energy of the excited polariton, while a negative (positive) imaginary part signals damping (amplification). At first order in $\omega_0/\kappa\ll 1$, the eigenfrequencies are given by
\begin{equation}
\omega_\pm=-i(\gamma_\downarrow - \gamma_\uparrow) \pm \omega_0\sqrt{\left(1-\frac{\bar{\eta}^2}{\eta_c^2}\right)\left(1-\frac{\Delta\eta^2}{\eta_c^2}\right)},
\end{equation}
as illustrated in~\cite{SI}. As the coupling $\bar{\eta}$ increases towards the critical point $\eta_c$, the phase transition occurs whenever any of $\Im[\omega_\pm]$ becomes positive~\cite{Dimer_2007, Nagy_2008, Eleuch_2013}. At large enough ratios $\Delta\eta/\bar{\eta}$, the damping rate $\gamma_\downarrow$ of the soft mode is dominant; this counteracts the coherent build-up of superradiance and stabilizes the normal phase.

We report in Fig.~\ref{fig:Fig3}(f) (blue shaded region) an estimate of the quasi-particle lifetime $\tau=-\left(2\Im[\omega_\pm]\right)^{-1}$ obtained from the spectrum of the eigenvalues. 
For this estimation we assume that the couplings $\bar{\eta},\Delta\eta$ are kept fixed at the end of the excitation pulse; this simplification provides a valid approximation for the measured lifetime where $\tau$ varies only slightly during the decay, i.e., as long as $\Delta\eta/\bar{\eta}\gtrsim 0.05$. For a closer comparison with the experimental data, we perform a numerical simulation of our experimental protocol including the excitation pulse [red shade in Fig.~\ref{fig:Fig3}(f)]. To account for collisional interactions and spin dephasing, the theoretical estimations include a phenomenological atomic damping (see~\cite{SI}), which we assume constant throughout the dynamics.\\

\hiddensection{Bistability and hysteresis}
\label{sec:Bistability}
\begin{figure}[thbp]
    \centering
    \includegraphics[width=1\columnwidth]{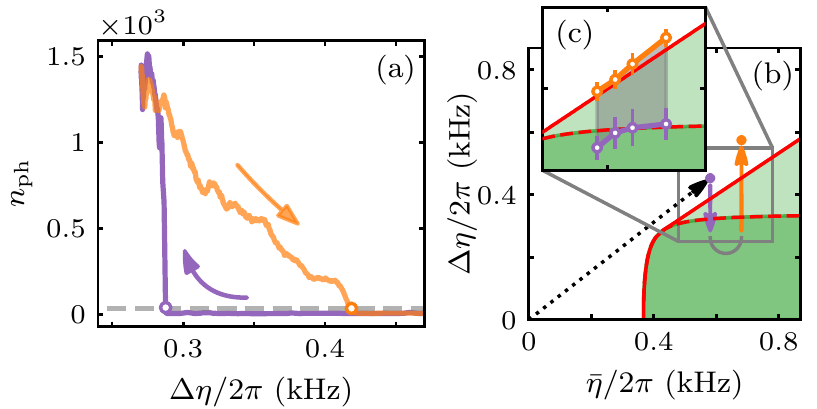}
    \caption{\textbf{Hysteresis at the boundary between the superradiant and the dissipation-stabilized normal phase}. (a) Time trace of the intra-cavity photons in a single typical experimental realization, when crossing the boundary between the superradiant and the dissipation-stabilized normal phase in opposite directions. (b) Corresponding trajectory in the parameter space $(\bar{\eta}, \Delta\eta)$. The hysteresis is measured at constant $\bar{\eta}$; an artificial offset along $\bar{\eta}$ has been introduced between the forward and backward path in the conceptual figure (b) for better visibility. The dashed arrow marks the preparation of the system within $6~$ms in the dissipation-stabilized normal phase, starting from zero coupling. Each (down, up) sweep of $\Delta\eta$ across the bistability region (purple, orange) is a $3~$ms-long s-shaped ramp. (c) Mapping the hysteresis region. Dots: phase boundary detected during the forward (purple) and backward (orange) path for different $\bar{\eta}$. The position of the boundaries are determined from the photon traces by setting a threshold of $36$ mean intra-cavity photons, as indicated in (a) with the gray line. The data points shown in (c) are mean values of $12$ to $18$ realizations. As a guide to the reader, in the background of (b,c), the analytical phase diagram highlights the region of stable normal phase (white), stable superradiant phase (dark green) and bistability (light green). The theoretical boundaries have been rescaled to the experimental data, with a single factor applied to both couplings~\cite{SI}. For this measurement, $N=1.10(8)\times10^5$, $\omega_0=2\pi\cdot40(5)~$kHz, $\Delta_c=-2\pi\cdot 3.0(5)~$MHz.}
        \label{fig:Fig4}
\end{figure}

We focus now on the boundary between the superradiant phase and the dissipation-stabilized normal phase observed at large imbalances $\Delta\eta/\bar{\eta}$. Previous theoretical works predicted an intermediate region where both phases are stable (bistability)~\cite{Soriente_2018, Parkins_2020}. Accordingly, the occurrence of a first-order phase transition is expected.
The bistability can be understood in terms of a competition between the coherent and dissipative processes described above. By increasing the relative imbalance $\Delta\eta/\bar{\eta}$, the damping $\gamma_\downarrow$ fosters the population of mode $\ket{0}$ and acts against the coherent coupling responsible for superradiance. 
In the limit of dominant dissipation, such damping makes the dissipation-stabilized normal phase the unique stable steady state. Conversely, in the regime of comparable coherent coupling and dissipation, the phase to which the system converges depends on its initial preparation. If the system is prepared in the normal phase, it remains stable because of the aforementioned damping. On the other hand, if the system is initially in the superradiant phase, the dissipative damping towards mode $\ket{0}$ is counteracted by the presence of a coherent intra-cavity field that contributes to keep mode $\ket{1}$ significantly populated. In other words, preparing the system in the organized, symmetry-broken superradiant phase makes it more rigid against transitions assisted by cavity dissipation.

To explore the boundary between the dissipation-stabilized normal phase and the superradiant phase, we initialize the system in the former phase by preparing the BEC and ramping up the coupling $\bar{\eta}$ above $\eta_c$ at fixed imbalance $\Delta\eta/\bar{\eta}=0.78$. From this initial state, the transition to the superradiant phase is crossed by reducing $\Delta\eta$ at constant $\bar{\eta}$. Then, within the same experimental realization, the direction of the $\Delta\eta$ sweep is reversed and the dissipation-stabilized normal phase is retrieved. The comparison between the forward and backward paths shows a hysteretic behavior at the phase boundary, in agreement with the expected discontinuous character of the transition, see Fig.~\ref{fig:Fig4}. By performing the hysteresis measurement at different coupling $\bar{\eta}$, an experimentally accessible region where the superradiant and the dissipation-stabilized normal phase are both stable is mapped out [Fig.~\ref{fig:Fig4}(c)]. We verified that implementing the parameter loop in the opposite direction also gives rise to hysteresis~\cite{SI}.\\

\hiddensection{Conclusion and Outlook}
We showed that varying the imbalance of co- and counter-rotating coupling terms between a quantum gas and a lossy optical cavity engenders a tunable competition between coherent and dissipative processes across the superradiant phase transition, leading to the emergence of a dissipation-stabilized phase and hysteresis. Combining the control over dissipative and coherent couplings with real-time access to the dynamics allowed us to identify the underlying microscopic processes determining the observed phase diagram. We note that, if the cavity dynamics is adiabatically traced out, our system maps to a driven-dissipative version of the anisotropic Lipkin-Meshkov-Glick (LMG) model~\cite{Lipkin_1965, Parkins_2008, Keeling_2019, Ribeiro_2019, Maghrebi_2020}-an important reference model for quantum magnetism that describes a many-body spin system with all-to-all interactions. 
Furthermore, interesting phenomena are expected near the boundary of the normal phase~\cite{Soriente_2020}; as visible in Figs.~\ref{fig:Fig3}(g,h), here the real parts of the eigenfrequencies of the low-lying polariton merge, and their imaginary part bifurcates such that one squeezes, while the other broadens. In the Dicke limit, this phenomenon is limited to the coupling interval between the bifurcation and the transition point to the superradiant phase, which is very narrow for typical experimental parameters~\cite{Dimer_2007}. Squeezing of fluctuations can be obtained on a much wider range of parameters at the boundary between the normal phase and the dissipation-stabilized normal phase, where dissipation suppresses the change of the system's steady state. Characterizing the fluctuations of the normal modes in this regime will shine light on the generation of squeezing in the vicinity of exceptional points in non-Hermitian systems~\cite{Reiter_2014, Weig_2020, Ramirez_2020}. Moreover, performing experiments with small atom or photon numbers would unveil effects beyond mean-field, such as quantum jumps in the bistability region, as predicted recently in Refs.~\cite{Parkins_2020Q, Chitra_2019}. Furthermore, combining our findings on prominent dissipative effects with the generation of cavity-mediated spin-orbit interaction~\cite{Deng_2014,Lev_2019} opens a way to study spin-orbit coupling in a dissipative setting.\\

\begin{acknowledgements}
We are grateful to M.~Landini and K.~Kroeger for contributions at early stages of the experiment, to T.~L.~Heugel, R.~Chitra, A.~Eichler, and F.~Piazza for fruitful discussions, and to A.~Frank for electronic support. F.~Ferri, R.R-M., F.~Finger, T.D, and T.E acknowledge funding from the Swiss National Science Foundation: project numbers 182650 and 175329 (NAQUAS QuantERA) and NCCR QSIT, from EU Horizon2020: ERCadvanced grant TransQ (project Number 742579). M.S. and O.Z. acknowledge financial support from the Swiss National Science Foundation through grants PP00P2\textunderscore1163818 and PP00P2\textunderscore190078, as well as the ETH Research Grant ETH-4517-1. \\

\end{acknowledgements}

\bibliographystyle{apsrev}
\let\oldaddcontentsline\addcontentsline
\renewcommand{\addcontentsline}[3]{}
\let\addcontentsline\oldaddcontentsline

\newpage
\beginsupplement
\onecolumngrid
\setstretch{1.25}
\begin{center}
\large
\textbf{Supplemental material}
\end{center}
\normalsize
\tableofcontents
\normalsize
\section{\textsc{Experimental details}}
\subsection{BEC preparation and Zeeman shift characterization}
We prepare a Bose-Einstein condensate (BEC) of $^{87}$Rb atoms in the $m_F=-1$ Zeeman sublevel in the $F=1$ hyperfine manifold of the $5^2\mathrm{S}_{1/2}$ electronic level, using radio-frequency assisted evaporation in a magnetic quadrupole trap. The atom cloud is optically transported and confined at the center of the cavity mode by an optical crossed dipole potential $V_\text{ext}$, with frequencies $[\omega_{hx},\omega_{hy},\omega_{hz}]=2\pi\cdot [220(3),24.6(8),170.1(3)]~$Hz.

We apply a magnetic field $\bold{B}=B_z\bold{e}_z$, with $B_z<0$. To measure the Zeeman shift $\omega_z$ between the sublevels $m_F=-1$ (high-energy level) and $m_F=0$ (low-energy level), we employ cavity-assisted Raman transitions. We illuminate the BEC with the transverse pump with frequency $\omega_r<\omega_c$. Close to the two-photon resonance $\omega_r\approx\omega_c-\omega_z$, a large fraction of the BEC is transferred to $m_F=0$  via two-photon processes involving absorption of photons from the red pump and emission into the cavity mode, with an increase of the kinetic energy by $2\hbar\omega_\text{rec}$. In this Raman process, light is scattered into the cavity at frequency $\tilde{\omega} = \omega_r+(\omega_z -2\omega_\text{rec})$, fulfilling energy conservation. We infer the Zeeman splitting $\omega_z$ by measuring the frequency $\tilde{\omega}$ of the photons leaking out of the cavity using a heterodyne detection scheme. 

\subsection{Transverse pump characterization}
The two transverse pumps (drives) are derived from the same laser source. Their frequencies $\omega_r$ and $\omega_b$ are independently adjusted by employing double-pass acoustic optical modulators (AOMs) in different optical paths and recombining the beams afterwards. A small fraction of the beam is split close to the vacuum chamber and overlapped with an optical local oscillator at frequency $\omega_\text{LO}$, also derived from the same laser, on an AC photodiode. The beat notes at frequencies $\omega_r-\omega_\text{LO}$ and $\omega_b-\omega_\text{LO}$ are electronically separated and employed for intensity stabilization of the individual pumps. The distance between the retro-reflecting mirror and the atomic cloud is carefully adjusted such that the two standing waves overlap maximally at the position of the atomic cloud. The lattice depth associated to each pump is calibrated by means of Kapitza-Dirac diffraction \cite{Gadway_2009s}. For all the measurements presented in the main text, we increase the transverse pump powers via ramps of the form $V_{r,b}(t)=\tilde{V}_{r,b}[3(t/t_\text{ramp})^2-2(t/t_\text{ramp})^3]^2$, where $\tilde{V}_{r,b}$ is the final lattice depth of the $\omega_{r,b}$ pump and $t_\text{ramp}$ is the duration of the ramp. 

\subsection{Heterodyne detection}
We monitor the photon field leaking out of the cavity by separating on a polarizing beam-splitter (PBS) the $y-$ and $z-$polarization modes, and detecting each of them with heterodyne setups. The detection branch for the $z-$polarization is used to produce the data discussed in this work. The auxiliary detection setup for the $y-$polarized mode is used to probe the cavity resonance at the end of each experimental cycle. 

In the heterodyne setup for the relevant $z-$polarized mode, the light field from the cavity is interfered with a local oscillator laser at frequency $\omega_\text{LO}$, and the high detection bandwidth of 250 MS/s allows for an all-digital demodulation of the beat note at $\omega-\omega_\text{LO}$ over a broad frequency range of $[0,125]$~MHz. In order to calibrate the photon number, we inject an on-resonance laser field into the empty cavity. We find the conversion factor between  the demodulated heterodyne signal and the mean intra-cavity photon number by measuring the power after the PBS with a powermeter, and using the knowledge on the losses of the cavity mirrors.

The complex intra-cavity field $\alpha(t)=X(t)-iY(t)$ is obtained from the quadratures $X(t)$ and $Y(t)$ after digital demodulation at frequency $\delta\omega_D=\bar{\omega}-\omega_\text{LO}$. The power spectral density (PSD) is calculated as PSD($f$)=$|\text{FFT}(\alpha)|^2(f)$ using a fast Fourier-transform of the form $\text{FFT}(\alpha)(f)=dt/\sqrt{N}\sum_i \alpha^*(t_i) e^{-i2\pi f t_i}$ \cite{Dogra_2019s}, where $t_i$ is the time of the $i^\text{th}$ step and $N$ is the total number of steps in the integration window. To construct the spectrograms, the traces are divided in time intervals of $T=150~\upmu$s with an overlap of $50\%$ between subsequent intervals. Finally, we calculate the photon number spectrograms as $\tilde{n}_\text{ph} =\text{PSD}(f)/T$.

\section{\textsc{Derivation of the Hamiltonian}}
\label{sec:A1}
In this Section, we derive the Hamiltonian in Eq.~\eqref{eq:Hamiltonian} from the closed-system dynamics of our spinor BEC coupled to the cavity.
\subsection{Single-atom Hamiltonian}
The Hamiltonian of a single atom coupled to the cavity mode reads
\begin{equation}
\hat{H}'_\text{1}=\hat{H}'_\text{at} + \hat{H}'_\text{cav} + \hat{H}'_\text{int}.
\label{eq:H_SP}
\end{equation}
In the dispersive regime of atom-light interaction~\cite{Goldman_2014s, Le_2013s}, the optically excited states of the atom can be adiabatically eliminated, and the Hamiltonian of the bare atom $\hat{H}'_\text{at}$ can be written in terms of the ground states levels $\ket{F,m_F}$ only:
\begin{equation}
\hat{H}'_\text{at} = \frac{\hat{\bold{p}}^2}{2m} + V_\text{ext}(\hat{\bold{x}}) + \sum_{F,m_F}\hbar\omega_{F,m_F}\ket{F,m_F}\bra{F,m_F},
\label{eqSI:H_at_full}
\end{equation}
where $\hat{\bold{p}}$ and $m$ are respectively the momentum and the mass of the atom, $V_\text{ext}(\hat{\bold{x}})$ is the trapping potential, which is kept fixed, and $\hbar\omega_{F,m_F}$ is the energy of the $\ket{F,m_F}$ atomic level, with $F$ denoting the hyperfine manifold, and $m_F$ the Zeeman sublevel. In our experiment, the $^{87}$Rb atoms are initialized in $\ket{F=1, m_F=-1}$, and near-resonant two-photon Raman transitions couple them to $\ket{F=1, m_F=0}$. Transitions to the $F=2$ manifold are off resonance by the hyperfine splitting $\omega_\mathrm{HF}=2\pi\cdot6.834~$GHz, and can be neglected. 
The internal dynamics of each atom can then be described in terms of the spin operator $\bold{\hat{F}}=(\hat{F}_x,\hat{F}_y,\hat{F}_z)^T$, with $F=1$. The energy difference between the Zeeman sublevels is determined by first- and second-order Zeeman shifts $\hbar\omega_z^{(1)}<0$ and $\hbar\omega_z^{(2)}>0$, such that Eq.~\eqref{eqSI:H_at_full} can be written as
\begin{equation}
\hat{H}'_\text{at} = \frac{\hat{\bold{p}}^2}{2m} + V_\text{ext}(\hat{\bold{x}}) + \hbar{\omega_z^{(1)}}\hat{F}_z + \hbar\omega_z^{(2)}\hat{F}_z^2.
\label{eqSI:H_at}
\end{equation}
The Hamiltonian of the bare cavity mode reads
\begin{equation}
\hat{H}'_\text{cav} = \hbar\omega_c \hat{a}^\dagger \hat{a},
\end{equation}
where the operator $\hat{a}^\dagger$ creates $z$-polarized photons in the TEM$_{00}$ cavity mode with resonance frequency $\omega_c$. In the dispersive regime at which we operate, the interaction between the light fields and the atom takes the form
\begin{equation}
\hat{H}'_\text{int} = \alpha_s\bold{\hat{E}}^{(+)}\cdot\bold{\hat{E}}^{(-)} - i\frac{\alpha_v}{2F}\left(\bold{\hat{E}}^{(+)}\times\bold{\hat{E}}^{(-)}\right)\cdot\bold{\hat{F}},
\label{eqSI:H_int}
\end{equation}
where $\bold{\hat{E}}^{(-)}$ ($\bold{\hat{E}}^{(+)}$) is the negative (positive) part of the total electric field at the position of the atom, with $\bold{\hat{E}}^{(+)}=(\bold{\hat{E}}^{(-)})^{\dagger}$, and $\alpha_s$, $\alpha_v$ are respectively the scalar and vectorial polarizability at the frequency of the driving lasers \cite{Goldman_2014s,Le_2013s,Landini_2018s}, with $\alpha_{s}<0$ and $\alpha_{v}>0$. In Eq.~\eqref{eqSI:H_int} we are neglecting an additional rank-2 tensor contribution to the polarizability, which is justified for $^{87}$Rb at the wavelength $\lambda=784.7~$nm at which we operate. 

We consider classical $y$-polarized transverse pump fields propagating along the $z$-direction at frequencies $\omega_{r,b}$, and a quantized cavity field. The negative part $\bold{\hat{E}}^{(-)}$ of the total electric field  is given by
\begin{equation}
\bold{\hat{E}}^{(-)}=\frac{E_{r}}{2}f_r(\hat{\bold{x}})\bold{e}_ye^{-i\omega_rt} + \frac{E_{b}}{2}f_b(\hat{\bold{x}})\bold{e}_ye^{-i\omega_bt} + E_{0}g(\hat{\bold{x}})\hat{a}\bold{e}_z,
\end{equation}
with unit vectors $\bold{e}_j$ ($j\in\{x,y,z\}$) and spatial mode profiles $f_r(\hat{\bold{x}})$, $f_b(\hat{\bold{x}})$, $g(\hat{\bold{x}})$. The two laser drives with amplitude $E_r$, $E_b$ originate from the same optical fiber, and their standing-wave modulations overlap in phase at the position of the trapping potential. Given the small frequency difference $\omega_b-\omega_r=2\pi\cdot96~$MHz, we can consider the same wavevector $k=\bar{\omega}/c$ for the two drives interacting with the atoms, with $\bar{\omega}=(\omega_b + \omega_r)/2$, and restrict to a single spatial profile $f_r(\hat{\bold{x}})=f_b(\hat{\bold{x}})=f(\hat{\bold{x}})=\exp[-2x^2/w_x^2-2y^2/w_y^2]\cos(kz)$. We also take $g(\hat{\bold{x}})=\exp[-2(y^2+z^2)/w_c^2]\cos(kx)$ for the cavity mode profile. The waist sizes of the modes are $[w_x, w_y, w_c]=[24,27,25]~\upmu$m. The amplitude of the cavity field per photon is defined by the frequency and volume of the mode, resulting in $E_{0}=403~$V/m.

We introduce the auxiliary Hamiltonian $\hat{H}_\text{rot}=\hbar\bar{\omega}\hat{a}^\dagger \hat{a} - \hbar\omega_z'\hat{F}_z$, and perform the unitary transformation $\hat{U}=\exp[\frac{i}{\hbar}\hat{H}_\text{rot} t]$, with $\omega_z'=(\omega_b-\omega_r)/2$. By making use of the rotating wave approximation, we obtain the time-independent Hamiltonian
\begin{equation}
\hat{H}_\text{1} = \hat{H}_\text{at} + \hat{H}_\text{cav} + \hat{H}_\text{s} + \hat{H}_\text{v},
\end{equation}
where
\begin{equation}
\hat{H}_\text{at} = \frac{\hat{\bold{p}}^2}{2m} + V_\text{ext}(\hat{\bold{x}}) + \hbar\delta_z\hat{F}_z + \hbar\omega_z^{(2)}\hat{F}_z^2,
\end{equation}
\begin{equation}
\hat{H}_\text{cav} =-\hbar\Delta_c \hat{a}^\dagger \hat{a},
\end{equation}
with cavity detuning $\Delta_c=\bar{\omega}-\omega_c$ and effective linear shift $\delta_z=\omega_z^{(1)}+\omega_z'$. The interaction part has a scalar and a vectorial contribution $\hat{H}_\text{s}$, $\hat{H}_\text{v}$ given by
\begin{equation}
\hat{H}_\text{s}=\frac{\alpha_s}{4}(E_b^2+E_r^2)f(\hat{\bold{x}})^2 + \alpha_s E_0^2\hat{a}^\dagger\hat{a}g(\hat{\bold{x}})^2,
\label{eqSI:H1_scalar}
\end{equation}
\begin{equation}
\hat{H}_\text{v} = \frac{\alpha_v}{8}E_0\left[\left(E_{b} + E_{r}\right)\left(\hat{a}+\hat{a}^\dagger\right)\hat{F}_x + \left(E_{b} - E_{r}\right)i\left(\hat{a}-\hat{a}^\dagger\right)\hat{F}_y\right]f(\hat{\bold{x}})g(\hat{\bold{x}})
\label{eqSI:H1_vectorial}
\end{equation}
respectively. Note that we applied a global rotation of the cavity field of the form $\hat{a}\rightarrow\hat{a}e^{i\pi/2}$. The first term in the scalar interaction $\hat{H}_s$  [cf. Eq.~\eqref{eqSI:H1_scalar}] describes the attractive potential created by the transverse drives, giving rise to a one-dimensional lattice along the $z$-direction, with on-axis depth $V_\text{TP}=-\alpha_s (E_b^2+E_r^2)/4$, and to an additional confinement along the $x-$ and $y-$direction. The second term in $\hat{H}_s$ is responsible for the dispersive shift of the cavity resonance and for a weak one-dimensional lattice potential along $x$-direction. We define the maximal dispersive frequency shift per atom as $U_0=\alpha_s E_0^2/\hbar$. The vectorial interaction $\hat{H}_v$ in Eq.~\eqref{eqSI:H1_vectorial} produces spin-changing Raman transition between the Zeeman sublevels of the $F=1$ manifold. The spin-changing terms $\hat{F}_{x}$, $\hat{F}_{y}$ are mediated by orthogonal quadratures of the cavity field and can be tuned by the sum or difference of the two pump fields, respectively.

\subsection{Many-body Hamiltonian}
We derive the Hamiltonian for the many-body system of $N$ atoms in a degenerate Bose gas using the second-quantization formalism. Using the single-atom results from the previous section, the many-body Hamiltonian can be written as
\begin{equation}
\hat{H}_\text{MB}=\hat{H}_\text{cav}+\int\hat{\Psi}^\dagger(\bold{x})\left(\hat{H}_\text{at}+\hat{H}_\text{s}+\hat{H}_\text{v}\right)\hat{\Psi}(\bold{x})d\bold{x},
\end{equation}
where the $\hat{\Psi}(\bold{x})$ is the spinor atomic field operator $\hat{\Psi}(\bold{x})=\left(\hat{\Psi}_{+1}(\bold{x}),\,\hat{\Psi}_{0}(\bold{x}),\,\hat{\Psi}_{-1}(\bold{x})\right)^T$,  fulfilling the bosonic commutation relations $\left[\hat{\Psi}_i(\bold{x}),\hat{\Psi}^\dagger_j(\bold{x'})\right]=\delta_{ij}\delta(\bold{x}-\bold{x}')$ and $\left[\hat{\Psi}_i(\bold{x}),\hat{\Psi}_j(\bold{x'})\right]=0$, with $i,j=+1,0,-1$. At this level, we neglect collisional interactions assuming low densities.

We set the half-frequency difference $\omega_z'$ between the drives close to the energy separation between levels $\ket{m_F=-1}$ and $\ket{m_F=0}$, i.e., $\omega_z'\approx\omega_z$, with $\omega_z=-\omega_z^{(1)}+\omega_z^{(2)}=2\pi\cdot48~$MHz.  Thus, spin-changing Raman transitions to $\ket{m_F=+1}$ are off-resonance by $\Delta_{+1}\approx2\omega_z^{(2)}=2\pi\cdot0.7~$MHz. The large detuning $\Delta_{+1}$ determines the fastest timescale of the atomic evolution. This allows to adiabatically eliminate the atomic operator $\hat{\Psi}_{+1}$, and restrict the dynamics to the two Zeeman sublevels with $m_F=0,-1$. The effect of the eliminated state leads to a sub-kHz energy shift of state $\ket{m_F=0}$, which we neglect. 

We map our system to an effective generalized Dicke model by further restricting the Hilbert space to two spin-momentum modes, in the same spirit as previous works \cite{Baumann_2010s, Lev_2018s}. In the normal phase, the BEC, prepared in $m_F=-1$, occupies the ground state of the total trapping potential, resulting from the combination of the external trap $V_\text{ext}$ and the attractive lattice potential $V_\text{TP}$ of the laser drives [cf. Eq~\eqref{eqSI:H1_scalar}]. We label this ground state as $\ket{0_a}$, with corresponding wave function  $\Phi_{0_a}(\bold{x})$. The cavity-mediated spin-changing interaction couples $\ket{0_a}$ to a density-modulated state $\ket{1_a}$ in $m_F=0$, with wave function $\Phi_{1_a}(\bold{x})=\mathcal{N}\Phi_{0_a}(\bold{x})\cos(kx)\cos(kz)$, with $\mathcal{N}$ being a normalization factor. Within this two-mode description, the spinor field operator takes the form $\hat{\Psi}(\bold{x})=\left(0,\,\Phi_{1_a}(\bold{x})\hat{c}_{1_a},\,\Phi_{0_a}(\bold{x})\hat{c}_{0_a}\right)^T$, where $\hat{c}_{0_a}$, $\hat{c}_{1_a}$ are bosonic annihilation operators for the respective atomic modes. The corresponding expression for the many-body Hamiltonian is
\begin{equation}
\hat{H}_\text{MB}=-\hbar\left[\Delta_c-N\mathcal{I}(V_\text{TP})U_0\right]\hat{a}^\dagger\hat{a}+\hbar\omega_0(V_\text{TP})\hat{J}_z
+ \frac{\alpha_v}{4\sqrt{2}}\mathcal{M}(V_\text{TP})E_0\left[(E_b+E_r)(\hat{a}+\hat{a}^\dagger)\hat{J}_x +i(E_b-E_r)(\hat{a}-\hat{a}^\dagger)\hat{J}_y\right],\label{eqSI:H_MB_complex}
\end{equation}
where we introduce collective pseudo-spin  $N/2-$operators $\hat{J}_x=(\hat{c}_{1_a}^\dagger\hat{c}_{0_a}+\hat{c}_{0_a}^\dagger\hat{c}_{1_a})/2$, $\hat{J}_y=(\hat{c}_{1_a}^\dagger\hat{c}_{0_a}-\hat{c}_{0_a}^\dagger\hat{c}_{1_a})/2i$, $\hat{J}_z=(\hat{c}_{1_a}^\dagger\hat{c}_{1_a}-\hat{c}_{0_a}^\dagger\hat{c}_{0_a})/2$. We indicate with $\hbar\omega_0(V_\text{TP})$ the energy difference between the bare atomic modes. 
The quantities $\mathcal{I}(V_\text{TP})$ and $\mathcal{M}(V_\text{TP})$ are overlap integrals defined by $\mathcal{I}(V_\text{TP})=\bra{0_a}g(\bold{x})^2\ket{0_a}/N$ and $\mathcal{M}(V_\text{TP})=\bra{0_a}f(\bold{x})g(\bold{x})\ket{1_a}/N$. In writing Eq.~\eqref{eqSI:H_MB_complex}, we neglected the dependence of the dispersive cavity shift on $\hat{J}_z$, which is a valid approximation whenever the system is not deep into the superradiant phase.

By considering $\hbar\omega_0(V_\text{TP})$ as a constant, and taking its value in the limit of small $V_\text{TP}$, i.e., $\hbar\omega_0=\hbar(\omega_z'-\omega_z+2\omega_\text{rec})$, we can write the Hamiltonian in Eq.~\eqref{eqSI:H_MB_complex} in terms of the parameters defined in the main text:
\begin{equation}
\hat{H}_\text{MB}=-\hbar\Delta_c\hat{a}^\dagger\hat{a}+\hbar\omega_0\hat{J}_z 
+ \hbar(\eta_b+\eta_r)(\hat{a}+\hat{a}^\dagger)\hat{J}_x+i\hbar(\eta_b-\eta_r)(\hat{a}-\hat{a}^\dagger)\hat{J}_y,
\label{eqSI:H_MB_simple}
\end{equation}
where we use the substitution $[\Delta_c-N\mathcal{I}(V_\text{TP})U_0]\rightarrow\Delta_c$, and define $\eta_{b(r)}=\frac{\alpha_v}{4\sqrt{2}}\mathcal{M}(V_\text{TP})E_0 E_{b(r)}$. Parametrizing Eq.~\eqref{eqSI:H_MB_simple} in terms of $\bar{\eta}=(\eta_b+\eta_r)/2$ and $\Delta\eta=(\eta_b-\eta_r)/2$ produces the Hamiltonian given in Eq.~\eqref{eq:Hamiltonian} of the main text. A slight rearrangement of terms gives 
\begin{equation}
\hat{H}_\text{MB}=-\hbar\Delta_c\hat{a}^\dagger\hat{a}+\hbar\omega_0\hat{J}_z
+ \hbar\eta_b(\hat{a}\hat{J}_+ +\hat{a}^\dagger\hat{J}_-) +\hbar\eta_r(\hat{a}^\dagger\hat{J}_++\hat{a}\hat{J}_-),\label{eqSI:H_MB_TC_ATC}
\end{equation}
with $\hat{J}_\pm=\hat{J}_x\pm i\hat{J}_y$. From Eq.~\eqref{eqSI:H_MB_TC_ATC}, it is apparent that the couplings $\eta_b,\eta_r$ tune the strength of the co- and counter-rotating term of the light-matter interaction, respectively. In the limit $\eta_r=0$, the Hamiltonian~\eqref{eqSI:H_MB_TC_ATC} reduces to the Tavis-Cummings model~\cite{Tavis_Cummings_1968s}.

\subsection{Mapping between Hamiltonian couplings and experimental parameters}
We describe the mapping between the measured power of the transverse pump beams and the Raman couplings $\eta_b$, $\eta_r$ introduced in the main text. The BEC is trapped in the combined potential of the harmonic confinement $V_\text{ext}(\bold{x})$ and of the attractive potential created by the transverse pumps $V_\text{TP}(\bold{x})=-V_\text{TP}f(\bold{x})^2$, which has contributions from the two drives, i.e. $V_\text{TP}=V_b+V_r$. We monitor the power of each drive and extract the corresponding value of $V_{b(r)}$ in real time as described in the previous section. 

To calculate the wave function $\Phi_{0_a}$, we consider spin-independent s-wave scattering and employ a Thomas-Fermi approximation for the interacting  BEC in the total trapping potential~\cite{Dalfovo_1999s}. Spin-changing collisions can be neglected due to the large second-order Zeeman shift $\omega_z^{(2)}=2\pi\cdot0.35~$MHz at which we operate~\cite{Stamper-Kurn_2013s}.  In addition, we treat the one-dimensional lattice created by the transverse pump in the limit of large depth, and approximate the lattice as a succession of independent harmonic traps. This is justified by the fact that, in our experiments, the superradiant phase transition occurs at large $V_\text{TP}\gtrsim25~\hbar\omega_\text{rec}$. We then calculate the three-dimensional overlap integrals $\mathcal{I}(V_\text{TP})$, $\mathcal{M}(V_\text{TP})$ by using the expressions of the mode functions $f(\bold{x})=\exp[-2x^2/w_x^2-2y^2/w_y^2]\cos(kz)$ and $g(\bold{x})=\exp[-2(y^2+z^2)/w_c^2]\cos(kx)$, with waist sizes $[w_x, w_y, w_c]=[24,27,25]~\upmu$m. The divergence of each Gaussian mode over the extension of the BEC is negligible. The Raman couplings $\eta_b$, $\eta_r$ are then found
\begin{equation}
\eta_{b(r)}=\frac{\mathcal{M}(V_\text{TP})}{2\sqrt{2}}\frac{\alpha_v}{\sgn{\alpha_s}\cdot\alpha_s}\sqrt{-\frac{U_0 V_{b(r)}}{\hbar}},
\label{eqSI:mapping_coupling}
\end{equation}
where $\alpha_v/\alpha_s=-0.928$ and $\sgn{\alpha_s}=-1$ at the wavelength of the laser drives.

The overlap integral $\mathcal{M}(V_\mathrm{TP})$ converges to $\mathcal{M}_\mathrm{max}=0.68$ at large lattice depths. In the regime $V_\text{TP}\gtrsim25~\hbar\omega_\text{rec}$ at which the superradiant phase transition occurs, $\mathcal{M}(V_\mathrm{TP})$ deviates from $\mathcal{M}_\mathrm{max}$ by less than $2\%$. For simplicity, we then assume $\mathcal{M}(V_\mathrm{TP})=\mathcal{M}_\mathrm{max}$ when applying the conversion in Eq.~\eqref{eqSI:mapping_coupling} throughout the paper.

\section{\textsc{Phase diagram}}\label{Phase_diagram}
\subsection{Analytical calculation of the steady state}
Starting from the  Hamiltonian in Eq.~\eqref{eq:Hamiltonian} of the main text, we consider dissipation due to photons leaking out from the cavity at rate $\kappa$ in the form of a Lindblad operator
\begin{align}
& \mathcal{L}[\hat{a}] = \kappa[2\hat{a} \hat{\rho} \hat{a}^\dagger - \{ \hat{a}^\dagger \hat{a} , \hat{\rho}\}]\,. 
\end{align}
We disregard the  spin decay rate due to the negligible spontaneous emission between Zeeman sublevels. At this level, we also neglect the role of spin dephasing, which would lead to a negligible shift of the phase boundaries for our experimental parameters. Using the master equation
\begin{equation}
\frac{d \hat{\rho}}{dt} = - \frac{i}{\hbar} \left[\hat{H}, \hat{\rho}\right] + \mathcal{L}[\hat{a}], 
\label{eq:ME}
\end{equation}
we obtain mean-field equations of motion (EOMs) of the form
\begin{align}
\begin{split}
\dv{t}\alpha & = i\Delta_c \alpha - i 2\sqrt{N}\bar{\eta} X - 2\sqrt{N}\Delta\eta Y - \kappa \alpha\,,\\
\dv{t}X & = -\omega_0 Y - 4\sqrt{N} \Delta\eta \alphaim Z \,, \\ 
\dv{t}Y & = \omega_0 X - 4\sqrt{N}\bar{\eta} \alphare Z \,, \\
\dv{t}Z & = 4 \sqrt{N}\bar{\eta} \alphare Y + 4\sqrt{N} \Delta\eta \alphaim X  \,, \label{eqSI:EOMs_Matteo} 
\end{split}
\end{align}
where the mean-field order parameters are $\langle \hat{a} \rangle = \sqrt{N} \alpha, \langle \hat{J}_x\rangle = NX$, $\langle \hat{J}_y\rangle = NY$ and $\langle \hat{J}_z\rangle = NZ$. Introducing the renormalized couplings $\bar{\eta}_N = \sqrt{N}\bar{\eta}\,, \Delta{\eta}_N = \sqrt{N}\Delta{\eta}$, and imposing the spin constraint $X^2+Y^2+Z^2 = \frac{1}{4}$, we can solve analytically for $\alpha,X,Y,Z$ in the steady state. The normal phase corresponds to the trivial steady-state solution $\alphare=\alphaim=X=Y=0$, $Z=-1/2$, with $\alphare=\Re[\alpha]$ and $\alphaim=\Im[\alpha]$. The non trivial solutions of Eq.~\eqref{eqSI:EOMs_Matteo}~\cite{Soriente_2018s} read
\begin{align}
\begin{split}
&\alphare = \pm\sqrt{c}\sqrt{\frac{2 a_2^2 b_1 + a_2 b_2^2 - 2 a_2 b_3 (a_1 + b_1) + 2 a_1 b_3^2 + \sgn{\Delta{\eta}_N - \bar{\eta}_N} a_2 |b_2| \sqrt{b_2^2 - 4 (a_1 - b_1)(a_2 - b_3) }}
	{2 (a_2^2 b_1^2 + a_1^2 b_3^2 + a_1 a_2 (b_2^2 - 2 b_1 b_3)) }}
\,, \\
&\alphaim = \frac{b_2^2 - \sgn{\Delta{\eta}_N - \bar{\eta}_N} |b_2|\sqrt{4 (b_1 - a_1)(a_2 - b_3) + b_2^2}}{2 b_2 (a_2 - b_3)}\alphare\,, \\
& X = -\frac{\Delta_c\alphare + \kappa\alphaim }{2\bar{\eta}_N}\,,\qquad Y = \frac{\Delta_c\alphaim - \kappa\alphare }{2\Delta{\eta}_N}\,,\qquad Z = \mathcal{A}\omega_0\,, \label{eq:steady_state}
\end{split}
\end{align}
with
\begin{align}
\begin{split}
& a_1=16\mathcal{A}^2\bar{\eta}_N^2\,,\qquad a_2=16\mathcal{A}^2\Delta{\eta}_N^2\,, \qquad b_1=\left(\kappa^2/\Delta{\eta}_N^2+\Delta_c^2/\bar{\eta}_N^2\right)/4\,, \\ &b_2=\kappa\Delta_c\left(1/\bar{\eta}_N^2-1/\Delta{\eta}_N^2\right)/2\,, \qquad b_3=\left(\kappa^2/\bar{\eta}_N^2+\Delta_c^2/\Delta{\eta}_N^2\right)/4\,, \qquad c=1/4 - \mathcal{A}^2 \omega_0^2\,,\\
& \mathcal{A} = -\frac{ \left(\bar{\eta}_N^2+\Delta{\eta}_N^2\right)\Delta_c - \sqrt{\left(\bar{\eta}_N^2-\Delta{\eta}_N^2\right)^2\Delta_c^2-4\kappa^2\bar{\eta}_N^2\Delta{\eta}_N^2}}{16\bar{\eta}_N^2\Delta{\eta}_N^2}.
\end{split}
\end{align}

\subsection{Analytic expressions for the phase boundaries}
We find an analytic expression for the slope of the dissipation-stabilized normal phase starting from the expression of $Z$ from Eq.~\eqref{eq:steady_state}:
\begin{equation}
Z = -\frac{ \left(\bar{\eta}_N^2+\Delta{\eta}_N^2\right)\Delta_c - \sqrt{\left(\bar{\eta}_N^2-\Delta{\eta}_N^2\right)^2\Delta_c^2-4\kappa^2\bar{\eta}_N^2\Delta{\eta}_N^2}}{16\bar{\eta}_N^2\Delta{\eta}_N^2}\omega_0.\label{eq:Z_constraint_explicit}
\end{equation}
Requiring $Z$ to be real results in the condition
\begin{equation}
\left(\bar{\eta}^2-\Delta{\eta}^2\right)^2\Delta_c^2-4\kappa^2\bar{\eta}^2\Delta{\eta}^2\ge 0.
\end{equation}
The equality provides the slope of the boundary between the superradiant phase and the dissipation-stabilized normal phase, cf.~Fig.~\ref{fig:Fig2} in the main text:
\begin{equation}
\label{eqSI:opening_line}
(\Delta\eta/\bar{\eta})_\mathrm{DSNP}=\kappa/\Delta_c\left(1-\sqrt{1+\Delta_c^2/\kappa^2}\right),
\end{equation}
for $\Delta_c<0$. Moreover, from Eq.~\eqref{eq:Z_constraint_explicit} we also obtain the stability boundary of the normal phase. Specifically, in the normal phase we set $Z=-1/2$ on the left hand side of Eq.~\eqref{eq:Z_constraint_explicit} and square both sides, we then solve for $\Delta{\eta}_N$ obtaining the following expression
\begin{equation}
	\Delta{\eta}_N = \frac{\sqrt{4 \bar{\eta}_N^2 |\Delta_c| \omega_0 - \omega_0^2 (\Delta_c^2+\kappa^2)}}{2\sqrt{4 \bar{\eta}_N^2- \omega_0 |\Delta_c|}},
\end{equation}
which allows us to find the boundaries of the bistability region shown in Figs.~\ref{fig:Fig4}(b,c) in the main text. 

\subsection{Numerical simulations of the mean-field dynamics}
In order to simulate the time evolution of the system, we numerically solve the semi-classical EOMs~\eqref{eqSI:EOMs_Matteo}. For this purpose, we use the MATLAB built-in `ode45' solver which is based on a Runge-Kutta (4,5) method~\cite{Shampine_1997s}. 
It employs variable time step sizes and the error tolerance in each step is constrained to $10^{-8}$. 
To sample the fluctuations on top of the mean-field observables and allow for a phase transition to take place, we assume an initially small photon field of the form $\alpha(t=0) = [\text{randn}(0,0.5)+i\cdot\text{randn}(0,0.5)]/\sqrt{N}$ with pseudo-random numbers $\text{randn}(0,0.5)$ sampled from a normal distribution with ($\mu,\sigma$)=(0,0.5). 
This assumption is compatible with an initial coherent vacuum state for the cavity field since $\langle\frac{\sqrt{N}(\alpha+\alpha^*)}{2}\rangle_\text{S}=0$ and $\text{var}_\text{S}\left(\frac{\sqrt{N}(\alpha+\alpha^*)}{2}\right)=1/4$, where $\langle\rangle_\text{S}$ and $\text{var}_\text{S}$ denote the average and variance over a sufficiently large number of samples $S$.

\begin{figure}
	\centering
	\includegraphics[width=0.7\columnwidth]{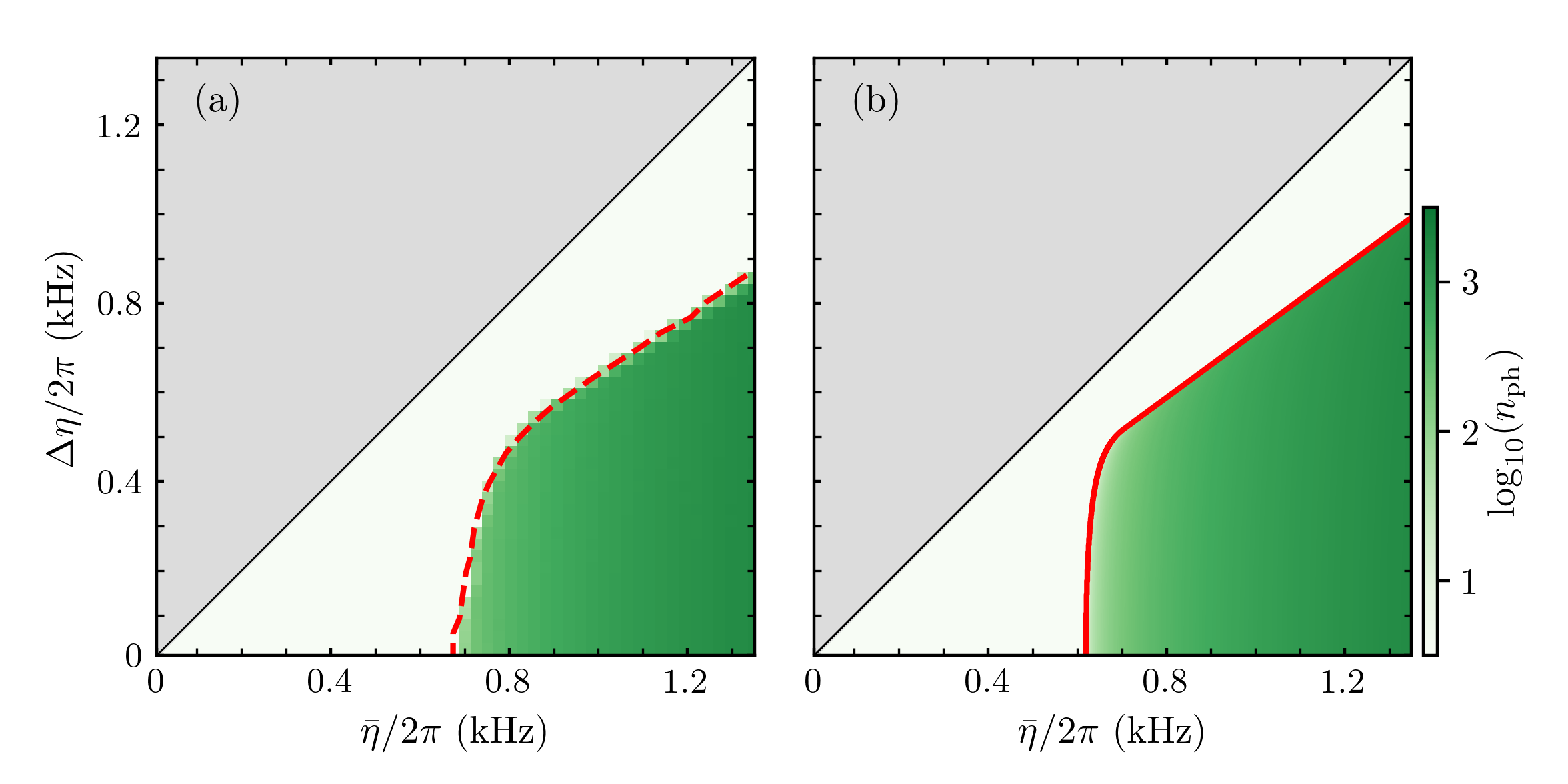}
	\caption{Theoretical phase diagrams from numerical mean-field simulations (a) and analytic steady-state calculations (b). For both methods, we consider $(\omega_0,\Delta_c,\kappa)=2\pi\cdot(44~\text{kHz},-4.0~\text{MHz}, 1.25~\text{MHz})$ and $N=1.28\times10^5$ atoms, to match the experimental parameters. For the numerics, we carry out 427 simulations at 61 different coupling ratios $\Delta\eta/\bar\eta$ and use s-shape coupling ramps with $t_\text{ramp}=10~$ms, as in the experiment. The boundary of the superradiant phase is marked in red.}
	\label{figSI:PhaseDiagramsTheory}
\end{figure}

We plot in Fig.~\ref{figSI:PhaseDiagramsTheory} the phase diagrams obtained from the numerical simulations and analytic steady-state calculations for the experimental parameters of Fig.~\ref{fig:Fig2} in the main text. The red lines in each plot indicate the boundaries of the superradiant phase assuming a threshold photon number of $n_\mathrm{ph,th}=5$. We attribute the small shift of the phase boundaries of the numerical simulations from the analytical results to residual non-adiabatic effects.

\subsection{Measurement of the phase diagram}
\subsubsection{Experimental protocol and data processing}
To measure the phase diagram, we prepare a BEC in $m_F=-1$ and ramp up the power of the driving lasers at constant cavity detuning $\Delta_c$. The calibrated heterodyne signal is used to construct photon number spectrograms $\tilde{n}_\text{ph}(f,t)$ as described in the next section. We integrate them in a narrow frequency range of $P=[\bar{\omega}/2\pi-2.5~\text{kHz},\,\bar{\omega}/2\pi+2.5~\text{kHz}]$ to obtain the photon traces $n_\mathrm{ph}(t)=\sum_{f\in P} \tilde{n}_\mathrm{ph}(f,t)$ of the cavity field at the frequency $\bar{\omega}$ characteristic of the superradiant phase.

The phase diagram in Fig.~\ref{fig:Fig2}(b) of the main text is obtained by combining measurements for $51$ different ratios $\Delta\eta/\bar{\eta}$, with $5$ realizations each. From each realization, we extract the coupling ramps $\bar{\eta}(t)$, $\Delta\eta(t)$ by monitoring in real time the power of the driving lasers, as well as the time trace of the mean cavity photons $n_\text{ph}(t)$ [cf. Fig.~\ref{fig:Fig2}(a) in the main text]. After parametrizing $n_\mathrm{ph}$ as a function of $\bar{\eta}$ and $\Delta\eta$, the colorplot in Fig.~\ref{fig:Fig2}(b) of the main text is obtained by averaging the different experimental realizations in the $(\bar{\eta},\Delta\eta)$ parameter space within squared bins with size $2\pi\cdot24~$Hz.

From each experimental realization, we extract the time $t_\text{th}$ at which the superradiant phase transition occurs by fitting $n_{\mathrm{ph}}(t)$ with a piecewise linear and power law function. The critical couplings are obtained as $(\bar{\eta}_\text{th},\Delta\eta_\text{th})=(\bar{\eta}(t_\text{th}),\Delta\eta(t_\text{th}))$. The dots in Fig.~\ref{fig:Fig2}(b) are the average of the critical couplings $(\bar{\eta}_\text{th},\Delta\eta_\text{th})$ from measurements taken with the same ratio $\Delta\eta/\bar{\eta}$. The errorbars are the corresponding standard error of the mean.

The experimental boundary between superradiant phase and the dissipation-stabilized normal phase [dashed line in Fig.~\ref{fig:Fig2}(b) in the main text] corresponds to the smallest value of $\Delta\eta/\bar{\eta}$ at which the phase transition is not observed in at least one of the experimental realizations, which we define as $(\Delta\eta/\bar{\eta})_\text{DSNP}$. The upper (lower) boundary of the gray shaded region around this line marks the smallest (largest) ratio $\Delta\eta/\bar{\eta}$ at which the phase transition is absent (present) in all realizations. These bounds provide an uncertainty to the slope $(\Delta\eta/\bar{\eta})_\text{DSNP}$.

\subsubsection{Phase diagrams for different cavity detunings}
\begin{figure}
	\centering
	\includegraphics[width=0.98\columnwidth]{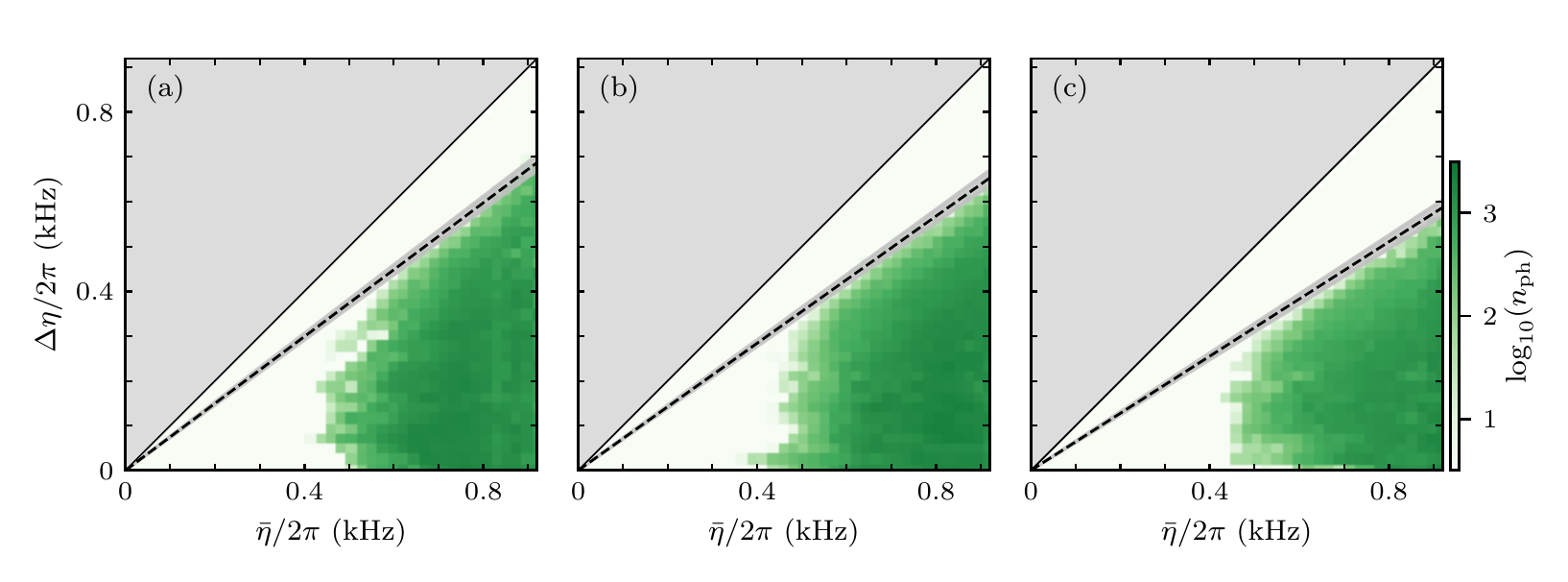}
	\caption{Phase diagrams for different cavity detunings $\Delta_c/2\pi = -5.0(2)~$MHz (a), $-4.0(2)~$MHz (b) and $-3.0(2)~$MHz (c). For these measurements, $N=1.28(8)\times 10^5$ and $\omega_0 =2\pi\cdot 44(2)~$kHz. Each of the phase diagrams is a collection of 200 to 350 individual realizations with different coupling imbalances $\Delta\eta/\bar{\eta}$. The slope of the dashed line corresponds to $(\Delta\eta/\bar{\eta})_\mathrm{DSNP}$ as described in the text. The shaded region around it marks the uncertainty on such slope.}
	\label{figSI:PhaseDiagramsDetuning}
\end{figure}

We record experimental phase diagrams for three different cavity detunings $\Delta_c$ and display the results in Fig.~\ref{figSI:PhaseDiagramsDetuning}. 
From each phase diagram, we extract the slope of the phase boundary $(\Delta\eta/\bar{\eta})_\mathrm{DSNP}$ and its uncertainty as discussed above.
The results of $(\Delta\eta/\bar{\eta})_\mathrm{DSNP}$ vs. $-\kappa/\Delta_c$ are shown in the inset of Fig.~\ref{fig:Fig2}(b) in the main text.

\subsubsection{Absence of additional self-organization processes}
The drive at frequency $\omega_r$, red-detuned from cavity resonance, may induce spurious atomic self-organization that does not involve a change of the $m_F$ state~\cite{Baumann_2010s}. If present, this process would be accompanied by the build-up of a coherent cavity field with the same polarization (along $y$) and frequency ($\omega_r$) as the driving laser. During our measurements, we continuously monitor such cavity field with the auxiliary heterodyne setup, and never observe any signal above noise. The suppression of spin-preserving self-organization is due to the large detuning $\vert\omega_r-\omega_c\vert$ and to the presence of the second drive at frequency $\omega_b$.

\section{\textsc{Excitation spectrum}}
\label{sec:A3}
\subsection{Spectrum of the open system}\label{subsec:spectrum_open_system}
To find the spectrum of the open system, we consider the system's EOMs~\eqref{eqSI:EOMs_Matteo} and expand the order parameters as $\alpha = \alpha_0 + \delta\alpha, X = X_0 + \delta X, Y = Y_0 + \delta Y, Z = Z_0 + \delta Z$ around the steady states $(\alpha_0, X_0, Y_0, Z_0)$, chosen to be either the normal or the superradiant phase. A similar treatment has been used in Refs.~\cite{Soriente_2018s, Soriente_2021s}. The linearized EOMs read
\begin{equation}
\dv{t}\begin{pmatrix}
	\delta \alphare \\
	\delta \alphaim \\
	\delta X \\
	\delta Y \\
	\delta Z \\
	\end{pmatrix} =
	\bold{M_0}
	\begin{pmatrix}
	\delta \alphare \\
	\delta \alphaim \\
	\delta X \\
	\delta Y \\
	\delta Z \\
	\end{pmatrix}\,,
\end{equation}
with
\begin{equation}
\bold{M_0}=\begin{pmatrix}
-\kappa & -\Delta_c & 0 & - 2\Delta{\eta}_N & 0 \\
\Delta_c & -\kappa & - 2\bar{\eta}_N & 0 & 0 \\
0 & -4 \Delta{\eta}_N Z_0 & - \Gamma_\phi & - \omega_0 & -4\Delta{\eta}_N \alphaim^0 \\
-4 \bar{\eta}_N Z_0 & 0 & \omega_0 & - \Gamma_\phi & -4\bar{\eta}_N \alphare^0 \\
4 \bar{\eta}_N Y_0 & 4 \Delta{\eta}_N X_0 & 4 \Delta{\eta}_N \alphaim^0 & 4 \bar{\eta}_N \alphare^0 & 0 \\
\end{pmatrix},
\label{eqSI:stability_matrix}
\end{equation}
where for completeness we phenomenologically included atomic dephasing at rate $\Gamma_\phi$. This damping term is compatible with a Lindblad term of the form $\mathcal{L}[\hat{J}_z]=\Gamma_\phi[2\hat{J}_z\hat{\rho}\hat{J}_z-\{\hat{J}_z\hat{J}_z,\hat{\rho}\}]$. From the diagonalization of the dynamical matrix in Eq.~\eqref{eqSI:stability_matrix}, we obtain the eigenfrequencies and eigenmodes of the system around the steady states. Since in our experiment $\vert\Delta_c\vert\gg\omega_0$, a clear separation between the photon-like and atom-like polariton modes exists. The polariton mode $\ket{1}$ discussed in the main text is the atom-like mode obtained by linearizing around the normal phase. The eigenfrequencies associated to this polariton mode are the eigenvalues $\omega_\pm$ presented in Figs.~\ref{fig:Fig3}(g,h). The $\omega_-$ branch corresponds to the annihilation of a particle in the unexcited mode $\ket{0}$ and the creation of a particle in the polariton mode $\ket{1}$. The $\omega_+$ branch corresponds to the opposite process.

\subsection{Polaritonic decay rates $\gamma_{\downarrow(\uparrow)}$}
The decay rates $\gamma_{\downarrow(\uparrow)}$ of the low-energy polariton modes can be derived from the diagonalization of the dynamical matrix in Eq.~\eqref{eqSI:stability_matrix}. Here, we derive the simplified analytical expression of $\gamma_{\downarrow(\uparrow)}$ given in Eq.~\eqref{eq:damping} in the main text, which results from a perturbative expansion of the eigenvalues of the system in the small parameter $\omega_0/\kappa\ll1$, which is well justified for our experiment. We do not consider spin dissipation and use the Keldysh action formulation \cite{Sieberer_2016s,Soriente_2020s}. First, we bosonize the spin using Holstein-Primakoff transformation, $\hat{S}^z =\hat{b}^{\dagger}\hat{b} -\frac{N}{2}$, $\hat{S}^+ =\sqrt{N -\hat{b}^{\dagger}\hat{b}}\,\hat{b}$, with $\hat{b}$ being a bosonic annihilation operator. We then write the Keldysh action in frequency domain and integrate out the cavity degree of freedom, obtaining the spin only action
\begin{equation}
S_\text{spin} = \int 
	\begin{pmatrix} \mathbf{\hat{b}}_c^* & \mathbf{\hat{b}}_q^*\end{pmatrix}
	\begin{pmatrix} 0 & [G_A^\text{spin}]^{-1} \\ [G_R^\text{spin}]^{-1} & D_K^\text{spin}\end{pmatrix}
	\begin{pmatrix} \mathbf{\hat{b}}_c \\ \mathbf{b}_q\end{pmatrix}\,,
\end{equation}
where the 4-component Nambu-spinor is given by $\mathbf{\hat{b}}_i = (\,\hat{b}_i(\omega)\,\hat{b}_i^*(-\omega)\,)$, $i=c,q$, and the inverse Green's functions and Keldysh component are
\begin{align}
	& [G_A^\text{spin}]^{-1} = ([G_R^\text{spin}]^{-1})^\dagger = \begin{pmatrix} \omega - \frac{N\eta_b^2}{i\kappa-\omega-\Delta_c} - \frac{N\eta_r^2}{-i\kappa+\omega-\Delta_c} - \omega_0 & -\frac{2N\eta_b \eta_r \Delta_c}{(\omega + i\kappa)^2 -\Delta_c^2} \\ 
	-\frac{2N\eta_b \eta_r \Delta_c}{(\omega + i\kappa)^2 -\Delta_c^2} & -\omega - \frac{N\eta_b^2}{-i\kappa+\omega-\Delta_c} - \frac{N\eta_r^2}{i\kappa-\omega-\Delta_c} - \omega_0 \end{pmatrix}, \label{eq:inverse_GA}\\
	& D_K^\text{spin} = -2i\kappa \begin{pmatrix} \frac{N\eta_b^2}{\kappa^2+(\omega+\Delta_c)^2} + \frac{N\eta_b^2}{\kappa^2+(\omega-\Delta_c)^2} & N\eta_b\eta_r(\frac{1}{\kappa^2+(\omega+\Delta_c)^2} + \frac{1}{\kappa^2+(\omega-\Delta_c)^2}) \\ 
	N\eta_b\eta_r(\frac{1}{\kappa^2+(\omega+\Delta_c)^2} + \frac{1}{\kappa^2+(\omega-\Delta_c)^2}) & \frac{N\eta_b^2}{\kappa^2+(\omega-\Delta_c)^2} + \frac{N\eta_b^2}{\kappa^2+(\omega+\Delta_c)^2} \end{pmatrix}\,.
\end{align}
We derive the eigenvalues as the zeros of the determinant of the inverse advanced Green's function in Eq.~\eqref{eq:inverse_GA},
\begin{equation}
\resizebox{.87\hsize}{!}{$\displaystyle{2N\eta_b^2 \left(-\Delta_c \omega_0-N\eta_r^2+i \kappa  \omega +\omega ^2\right)+\left[N\eta_r^2+(\omega_0-\omega ) (-\Delta_c+i \kappa +\omega )\right]\left[N\eta_r^2+(\omega +\omega_0)(-\Delta_c-i \kappa -\omega )\right]+N^2\eta_b^4=0\,.}$}
	\label{eq:determinant}
\end{equation}
We perform a first order expansion in $\omega_0/\kappa\ll1$ by approximating $\omega + i\kappa \approx i\kappa$ and solve Eq.~\eqref{eq:determinant} for $\omega$. We obtain
\begin{align}
	\omega_\pm & \approx iN\frac{\kappa}{\kappa^2+\Delta_c^2}(\eta_b^2 - \eta_r^2) \pm \frac{\sqrt{(-4N\Delta\eta^2\Delta_c+(\kappa^2+\Delta_c^2)\omega_0)(-4N\bar{\eta}^2\Delta_c+(\kappa^2+\Delta_c^2)\omega_0)}}{\kappa^2+\Delta_c^2}, \\
	& = -i(\gamma_\downarrow - \gamma_\uparrow) \pm \omega_0\sqrt{\left(1-\frac{\bar{\eta}^2}{\eta_c^2}\right)\left(1-\frac{\Delta\eta^2}{\eta_c^2}\right)}.
	\label{eqSI:eigenvalues_approx}
\end{align}
We note that this result can be obtained also from a linear analysis after adiabatic elimination of the cavity field. 
The rates $\gamma_{\downarrow(\uparrow)}$ describe the dissipative damping (amplification) of the polariton mode $\ket{1}$, as discussed in the main text. They can be re-written in the form
\begin{equation}
\gamma_{\downarrow(\uparrow)}=N\eta^2_{b(r)}\rho(\bar{\omega}),
\label{eqSI:gamma_approx}
\end{equation}
where $\rho(\bar{\omega})=\kappa/[(\bar{\omega}-\omega_c)^2+\kappa^2]$ is the density of states of the cavity at the frequency $\bar{\omega}$ of the cavity field, cf.~Fig.~\ref{fig:Fig1} in the main text. The expression~\eqref{eqSI:gamma_approx} indicates that the mechanism at the origin of damping (amplification) $\gamma_{\downarrow(\uparrow)}$ is the scattering of photons from a single drive with strength $\eta_{b(r)}$ into the bath of vacuum modes provided by the cavity, accompanied by a transfer of population from mode $\ket{1}(\ket{0})$ to mode $\ket{0}(\ket{1})$.  

The expression in Eq.~\eqref{eqSI:gamma_approx} is valid in the limit $\omega_0\ll\kappa$. A more accurate estimation for $\gamma_{\downarrow(\uparrow)}$ can be obtained using Fermi's golden rule~\cite{Demler_2020s}, and the limit in which only a single drive is present, i.e., $\eta_b=0$ or $\eta_r=0$. The result is 
\begin{equation}
\gamma_{\downarrow(\uparrow)} = N\eta^2_{b(r)}\rho(\tilde{\omega}_{+(-)}),
\label{eqSI:gamma_Fermi_golden_rule}
\end{equation}
with $\tilde{\omega}_\pm=\bar{\omega}\pm\omega_0$. The frequency of the field scattered into the cavity by each drive deviates from $\bar{\omega}$ by $\pm\omega_0$, according to energy conservation [see also Fig.~\ref{fig:Fig1}(c) in the main text for a schematic visualization]. 

The correction obtained by using Fermi's golden rule becomes particularly relevant near the Dicke limit $\eta_b=\eta_r$. Without this correction, the rates $\gamma_{\downarrow(\uparrow)}$ compensate each other [cf.~Eq.~\eqref{eqSI:gamma_approx}], and the imaginary part of Eq.~\eqref{eqSI:eigenvalues_approx} is zero for all $\bar{\eta}<\bar{\eta}_c$, resulting in a vanishing damping rate. The result obtained with Fermi's golden rule allows to account for higher orders of $\omega_0/\kappa$, leading to a nonzero damping. By plugging  Eq.~\eqref{eqSI:gamma_Fermi_golden_rule} into the expression for the eigenvalues Eq.~\eqref{eqSI:eigenvalues_approx}, we find that, in the Dicke limit, the effective damping rate of the polariton mode $\ket{1}$ is 
\begin{equation}
\gamma_\downarrow-\gamma_\uparrow=-4N\eta_b^2\frac{\kappa\Delta_c\omega_0}{(\Delta_c^2+\kappa^2)^2},
\end{equation}
where we again made use of $\omega_0\ll\kappa$. This result is in agreement with previous derivations~\cite{Keeling_2012s} and provides insights into the physical origin of a finite effective polariton damping in the driven-dissipative Dicke model.

We point out that the theoretical results used for the analysis of excitations (Fig.~\ref{fig:Fig3} in the main text) have been obtained by exact diagonalization of the dynamical matrix~\eqref{eqSI:stability_matrix}, and therefore they do not suffer from truncation effects.

\subsection{Probing excitations}
\subsubsection{Experimental protocol}
To measure the evolution of the excitation spectra, we prepare a BEC of $N=9.6(4)\times10^4$ atoms in $m_F=-1$ and ramp up the coupling strengths $\eta_{r,b}(t)$ within $t_\text{ramp}=9.1~$ms. The experimental parameters for these measurements are $\omega_0=2\pi\cdot 48(4)~$kHz and $\Delta_c=-2\pi\cdot5.8(1)~$MHz. While ramping up the coupling, we inject an excitation field through the cavity between $t\in[3.0,4.0]~$ms. The amplitude of the excitation field corresponds to $7.2(1)$ intra-cavity photons; its frequency is chosen to be close to the polariton resonance $\bar{\omega}+\omega_0$. By this method, we typically transfer $<10 \% $ of the atomic population in the excited polariton mode $\ket{1}$. After the end of the excitation pulse, the polariton mode evolves freely according to the dynamics of the open system. We monitor this free evolution in real time via the spectrum of the associated photon field.

\subsubsection{Relation between the polariton dynamics and the cavity spectrogram}
Here, we show how the dynamics of the polaritonic excitation can be derived from the associated cavity field, detected with a heterodyne setup, as done in Sec.~IV of the main text. Since the population in mode $\ket{1}$ prepared by the excitation pulse is small, we can linearize the mean-field dynamics of the system around the normal phase. By substituting $(\alpha_\mathrm{Re}, \alpha_\mathrm{Im}, X_0, Y_0, Z_0)=(0,0,0,0,-1/2)$ in Eq.~\eqref{eqSI:stability_matrix} and linearizing the pseudo-spin conservation $\delta Z=-(X_0\delta X+Y_0\delta Y)/Z_0=0$, we obtain a 4x4 stability matrix $\textbf{M}$ in the basis $(\delta \alphare,\delta \alphaim,\delta X,\delta Y)^T$. Diagonalization of this matrix leads to
\begin{equation}
\textbf{D} = \textbf{S}^{-1}\textbf{M}\textbf{S}= 
\begin{pmatrix}
-i\Delta_+ & 0 & 0 & 0 \\
0 & -i\Delta_- & 0 & 0 \\
0 & 0 & -i\omega_+ & 0 \\
0 & 0 & 0 & -i\omega_- \\
\end{pmatrix},
\label{eqSI:diag_matrix_NP}
\end{equation}
where $\Delta_+=-\Delta_-^*$ and $\omega_+=-\omega_-^*$ because $\textbf{M}$ has real coefficients. 
The pairs $\Delta_\pm$, $\omega_\pm$ correspond to the photon-like and atom-like polariton modes, respectively, with $\vert\Re[ \Delta_\pm]\vert\approx\vert\Delta_c\vert$ and $\vert\Re[\omega_\pm]\vert\leq\omega_0$ \cite{Domokos_2011s}. 
The eigenfrequencies depend on the couplings $\bar{\eta}, \Delta\eta$, which are time-dependent in our experimental protocol. The coupling sweeps are however slow enough to allow the system to evolve adiabatically,
and the polariton modes evolve independently of each other. 
By decomposing into polariton modes, the evolution of the cavity field quadratures can be written as
\begin{equation}
\alphare(t)=\delta\alphare(t)=\sum_{j=\pm}c_{pj}e^{i\Delta_jt}+\sum_{j=\pm}c_{aj}e^{i\omega_jt}\,,
\end{equation}
where $c_{lj}$ are complex coefficients, with $c_{l+}=c_{l-}^*$ and $l=p,a$ denoting photon and atom, respectively. Due to the large separation $\vert\Delta_c\vert\gg\omega_0$, the photon-like mode cannot be excited by the external pulse on resonance with the atom-like mode, and can be neglected. We thus find
\begin{equation}
\alphare(t)=c_0\cos(\Re[\omega_+] t+\phi_0)e^{\Im[\omega_+] t}.\label{eqSI:field_excitations}
\end{equation}
The initial conditions $c_0,\phi_0$ are determined by the externally induced excitation process. 
An analogous expression holds for $\alphaim$, allowing to directly relate the real and imaginary part of the polariton frequency to the measured cavity output. Specifically, from Eq.~\eqref{eqSI:field_excitations} we find that, since $\Im[\omega_\pm]<0$, the amplitude of the field $n_\mathrm{ph}$ decays as $n_\mathrm{ph}\propto e^{2\Im[\omega_\pm]t}$, with a corresponding $1/e$-decay time $\tau=-(2\Im[\omega_\pm]t)^{-1}$, which is the result used in Sec.~IV of the main text.\\

\subsubsection{Data processing and comparison to theory}
The experimental values of the excitations lifetime shown in Fig.~\ref{fig:Fig3}(f) of the main text are extracted from the heterodyne measurement of the cavity output in the following way. We consider the photon number spectrogram $\tilde{n}_{\mathrm{ph}}(\omega, t)$ and set $t=0$ to the end of the excitation pulse. 
We first extract the time $t_c$ at which the superradiant phase transition occurs by integrating $\tilde{n}_{\mathrm{ph}}(\omega, t)$ on a frequency interval $-\omega_{\mathrm{lim}}\leq\omega-\bar{\omega}\leq\omega_{\mathrm{lim}}$ with $\omega_{\mathrm{lim}}=2\pi\cdot 10~\text{kHz}\approx0.2\omega_0$ and setting a transition threshold of $100$ intra-cavity photons. Such threshold is large enough to capture only the coherent field in the superradiant phase, but still small enough to detect reliably the critical point. For large ratios $\Delta\eta/\bar{\eta}$ leading to the dissipation-stabilized normal phase, $t_c$ is taken as the average of the transition times extracted where the superradiant phase builds up. To study the time evolution of the polaritonic excitations, we integrate $\tilde{n}_{\mathrm{ph}}(\omega, t)$ on a larger frequency range $\omega_{\mathrm{lim}}=2\pi\cdot 150~\text{kHz}\sim3\omega_0$ to get the photon trace $n_\mathrm{ph}(t)$, with a time resolution of $10~\upmu$s. We extract the lifetime $\tau$ from the cumulative signal $s(t)=\int_0^t n_\mathrm{ph}(t')dt'$. For $n_\mathrm{ph}\propto e^{-t/\tau}$, $s(t)$ takes the form $s(t)=s_\mathrm{max}(1-e^{-t/\tau})$. Since for most of the data points the lifetime is significantly shorter than the transition time $t_c$, we determine $\tau$ by the time at which $s(t)$ has reached a fraction $(1-e^{-1})$ of its maximum below $t_c$.  The results of this estimation are in agreement with the ones obtained from a fit of $s(t)$, but more robust especially for weak signals. As an exception, for the single dataset at $\Delta\eta/\bar{\eta}=0$ the condition $\tau\ll t_c$ is not fulfilled, and $\tau$ is extracted from a fit of $s(t)$ with the exponential model.
We neglect the experimental realizations in which the atomic response during the excitation pulse is below the noise level. The data shown in Fig.~\ref{fig:Fig3}(f) of the main text are averaged values of $\tau$ over 10 to 25 realizations, with the error bar representing the maximum between standard error of the mean and the time resolution of the photon trace.\\

We compare the values of $\tau$ extracted experimentally with the theoretical expectations from the excitation eigenfrequencies of the system. 
According to the description given above, we expect that $n_\mathrm{ph}(t)\propto e^{2\Im[\omega_\pm]t}$, which provides an analytical estimation of the lifetime $\tau_\mathrm{an}=-\left(2\Im[\omega_\pm]\right)^{-1}$. 
If the couplings vary in time, the decay of the photon number $n_\mathrm{ph}(t)$ is in general non-exponential. 
However, a meaningful estimation for the measured lifetime $\tau$ at large enough imbalance ratio $\Delta\eta/\bar{\eta}\gtrsim 0.05$ is provided by the value $\tau_\mathrm{an}$ obtained for the instantaneous couplings $\bar{\eta}, \Delta\eta$ just after the excitation pulse.\footnote{Using the knowledge of the coupling ramps $\bar{\eta}(t) $, $\Delta\eta(t)$, we parametrize $\tau_\mathrm{an}$ as a function of time $\tau_\mathrm{an}=\tau_\mathrm{an}(t)$. We compare the lifetime after the excitation pulse $\tau_\mathrm{an}(t=0)$ with its variation $\delta\tau$ within the decay time of the excitations, estimated as $\delta\tau=\frac{\tau_\mathrm{an}(0)-\tau_\mathrm{an}(t_b)}{t_b}\tau_\mathrm{an}$, where $t_b$ is the time corresponding to the bifurcation. For $\Gamma_\phi=0$ and large enough imbalance ratios $\Delta\eta/\bar{\eta}>0.05$, we get $\delta\tau/\tau_\mathrm{an}<0.1$. Closer to the Dicke limit, the estimation is sensitive to the assumption on the atomic dephasing rate $\Gamma_\phi$, and the condition $\delta\tau/\tau_\mathrm{an}\ll 1$ is only fulfilled for large values of $\Gamma_\phi$.}

To account for processes leading to dephasing of the individual atomic spins, such as collisions, we introduce a phenomenological atomic dephasing rate $\Gamma_\phi$, as described in \ref{subsec:spectrum_open_system}. In Fig.~\ref{fig:Fig3}(f) of the main text, the blue shaded region shows the lifetime estimated from the eigenvalues, and assuming $\Gamma_\phi=0$ (upper bound) and $\Gamma_\phi=2\pi\cdot500~$Hz (lower bound, corresponding to the estimated collision rate in the total trapping potential). 

For a closer comparison to the experiment, a numerical simulation including time varying coupling is performed, using the method described below in~\ref{subsec:simulations_excitations} [red shaded region in Fig.~\ref{fig:Fig3}(f) of the main text]. 
The shaded region includes results of simulations performed for different initial phase $\phi_{\mathrm{probe}}\in[0, 2\pi)$ of the excitation drive, which we do not control in the experiment, and an atomic dephasing rate $\Gamma_\phi$ varying in the same interval described in the previous paragraph.

\subsection{Numerical simulations of the probing method}\label{subsec:simulations_excitations}
We simulate numerically the experimental protocol that we implemented to probe the excitation spectrum of our system, and described in Sec.~IV of the main text. For this purpose, we extend the theoretical model and incorporate an additional intra-cavity probe beam. We consider a classical $z$-polarized electric field propagating along the cavity axis
\begin{small}
\begin{equation}
E_{\mathrm{probe}}(t,x) = \tilde{E}_{\mathrm{probe}}\sqrt{n_{\mathrm{probe}}(t)}\cos(k_\text{rec} x)\text{e}^{-i\omega_\text{probe}t-i\phi_{\mathrm{probe}}},
\end{equation}
\end{small}
with $\omega_\text{probe}=\bar{\omega}+\omega_0+\delta_{\mathrm{probe}}$. Here, $n_{\mathrm{probe}}(t)$, $\phi_{\mathrm{probe}}$ and $\delta_{\mathrm{probe}}$ are the average intra-cavity photon number, relative phase and detuning with respect to the cavity field associated with the polariton branch $\omega_+$ at low couplings ($\bar{\omega}+\omega_0$). Moreover, $\tilde{E}_\text{probe}$ is the electric field per photon in this beam.

Following an analogous approach to the derivation of the Hamiltonian $\hat{H}$ in Sec.~\ref{sec:A1}, we obtain a time-dependent many-body Hamiltonian describing the interaction of the light-matter system with the probe field:
\begin{equation}
\hat{H}_\text{exc} = \hat{H}_\text{MB}
+4\hbar\bar{\eta}\sqrt{n_{\mathrm{probe}}(t)}\sin\left[(\omega_0+\delta_{\mathrm{probe}})t+\phi_{\mathrm{probe}}\right] \hat{J}_x
+4\hbar\Delta\eta\sqrt{n_{\mathrm{probe}}(t)}\cos\left[(\omega_0+\delta_{\mathrm{probe}})t+\phi_{\mathrm{probe}}\right] \hat{J}_y.\label{eqSI:HamiltonianExc}
\end{equation}
The probe beam drives the atomic coherences $\hat{J}_{x,y}$, similar to cavity-enhanced Bragg spectroscopy techniques \cite{Mottl_2012s}. Hence, we expect to coherently transfer non-negligible atomic populations to the excited state if we approach the low-coupling two-photon resonance $\delta_{\mathrm{probe}}\rightarrow 0$. 
\begin{figure}
	\centering
	\includegraphics[width=0.6\columnwidth]{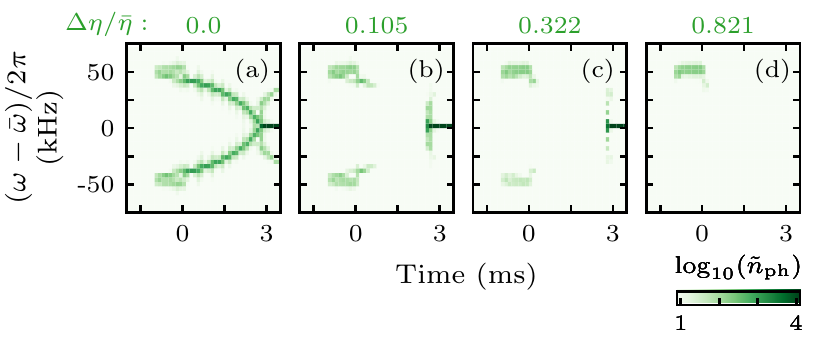}
	\caption{Spectrograms of the cavity field from numerical simulations of our experimental protocol to probe excitations, cf.~Fig.~\ref{fig:Fig3} in the main text. We consider the coupling imbalances $\Delta\eta/\bar{\eta}=0$(a), 0.105(b), 0.322(c) and 0.821(d). For the numerical simulation, we choose $(\omega_0,\omega,\kappa,\Gamma)=2\pi\cdot(48~\text{kHz},5.8~\text{MHz},1.25~\text{MHz},100~\text{Hz})$ and $N=9.6\times10^4$ atoms. The couplings are increased via an s-shaped ramp within 9.1 ms to $\bar{\eta} < 2\pi\cdot 1.16~\text{kHz}$ at fixed ratio $\Delta\eta/\bar{\eta}$. Moreover, a blue detuned probe with $\delta_{\mathrm{probe}} = 2\pi\cdot 2~\text{kHz}$, $n_{\mathrm{probe}}=7.2$ photons and $\phi_{\mathrm{probe}}=0$ illuminates the system between $-1~\text{ms} < t <0~\text{ms}$.}
	\label{figSI:NummericsExcitations}
\end{figure} 

We derive mean-field equations of motion from the Hamiltonian in Eq.~\eqref{eqSI:HamiltonianExc} and numerically solve them to simulate the experimental conditions of our excitation probing method, cf. Fig.~\ref{fig:Fig3} in the main text. We plot the resulting spectrograms of the PSD in Fig.~\ref{figSI:NummericsExcitations}. The coupling imbalances $\Delta\eta/\bar{\eta}$ and the parameters $(\omega_0,\Delta_c,\kappa,N)$ are chosen in accordance to the experimental observations reported in the main text [Figs.~\ref{fig:Fig3}(b)-(e)]. Depending on the choice of $\Delta\eta/\bar{\eta}$ we transfer between $8\%$ and $12\%$ of the atoms to the excited state at the end of the probe pulse. The results from the numerical simulations are in good agreement with the experimental results at different coupling imbalances: while in the Dicke limit the excitations are long-lived and a complete mode softening towards the superradiant phase is observed, the polaritonic excitations damp faster for increasingly large coupling imbalances, as observed in the experiment. 

\section{\textsc{Hysteresis measurement}}
\subsection{Data Processing}
For every single hysteresis measurement, we fix the coupling strength $\bar{\eta}$ and record the photon number $n_\text{ph}$ extracted from the heterodyne detector as a function of the other coupling strength $\Delta\eta$. To further process the raw data, we smoothen it by applying a moving average over 51 subsequent points. Subsequently, we define a threshold photon number for the detection of a stable superradiant phase.
 
In order to extract the threshold of the superradiant region, we set a threshold of $36$ mean photons, which is $5$ times above the noise level. Around this value, the width of the hysteresis is independent on the choice of the threshold. We determine and compare the critical coupling strength for the forward (backward) path, as shown in purple (orange), in Fig.~4 of the main text and in Fig~\ref{figSI:Hysteresis} below. The critical couplings are used to map out the hysteresis region for different coupling strengths $\bar{\eta}$. 
For every data point we take on average 15 measurements, with at least 12 and at most 18 repetitions.
\begin{figure}[thbp]
	\centering
	\includegraphics[width=0.5\columnwidth]{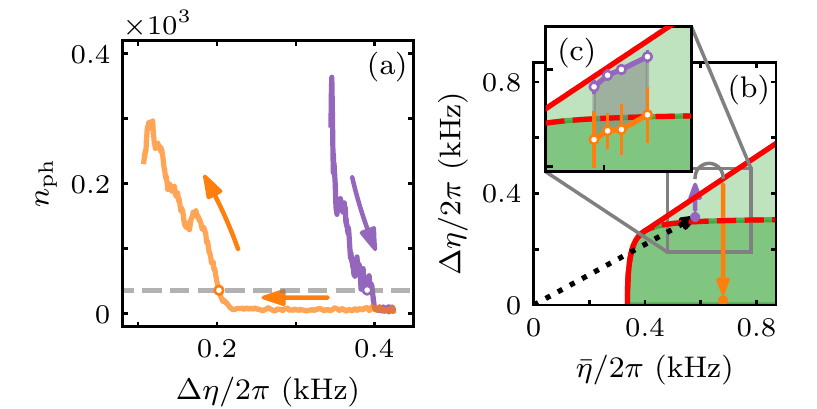}
	\caption{Hysteresis at the boundary between the superradiant phase and the dissipation stabilized normal phase. The hysteresis is measured starting in the superradiant phase, and performing a loop of the coupling $\Delta\eta$ in opposite direction as the one shown in the main text. (a) Exemplary time trace of the mean photon number $n_\mathrm{ph}$ during the loop. (b) Corresponding trajectory in the parameter space $(\bar{\eta},\Delta\eta)$. An artificial offset in $\bar{\eta}$ has been introduced between the forward and backward paths for better visibility. (c) Boundaries of the normal phase detected during the forward (purple) and backward (orange) path for different $\bar{\eta}$. The position of the boundaries is determined from
the photon traces by setting a threshold of 36 mean intra-cavity photons, as indicated with a gray line in (a). As  guide to the reader, in the background of (b,c), the phase diagram from analytical calculations shows the region of stable normal phase (white), stable superradiant phase (dark green) and bistability
(light green). The theoretical boundaries have been rescaled to the experimental data, with a single factor applied to both couplings. This scaling factor is chosen to overlay the theoretical phase boundary between the superradiant and bistability regions (dashed red line) and the corresponding experimental datapoint with the largest coupling $\bar{\eta}$.  For these measurements, we employ the following experimental parameters $N=1.10(8)\times 10^5$, $\Delta_c =- 2\pi\cdot 3.0(5)~$MHz and  $\omega_0 = 2\pi\cdot 40(5)~$kHz.} 
	\label{figSI:Hysteresis}
\end{figure} 

\subsection{Hysteresis loops in opposite directions}
Our hysteresis measurement is potentially sensitive to atom loss and heating during the experimental protocol. These processes affect the collective atom-cavity coupling, and consequently shift the stability boundaries of the different phases. To ensure that the measured bistability region is not substantially biased by a variation of the collective coupling due to these effects, we complement the measurement shown in Fig.~\ref{fig:Fig4} of the main text with the result of a hysteresis loop performed in the opposite direction, see Fig.~\ref{figSI:Hysteresis}. 

Hysteresis is observed also in this second measurement protocol, confirming that the effect of atom loss and heating is not substantial. 

\bibliographystyle{apsrev}

\end{document}